\documentclass{article}

\usepackage{subfigure,graphicx,dsfont,amsmath,amsthm,amssymb,psfrag,a4wide}

\newtheorem{lemma}{Lemma}

\newtheorem{theorem}{Theorem}
\newtheorem{proposition}{Proposition}
\newtheorem{corollary}{Corollary}

\newcommand{\abc}{a}

\newcommand{\load}{\nu}
\newcommand{\fsolv}{\load^*}
\newcommand{\ar}{r}
\newcommand{\avrth}{\bar{\theta}}
\newcommand{\indi}[1]{{\rm I}_{\{#1\}}}

\psfrag{index}{$i$}
\psfrag{theta_i}{\small{$\theta_i$}}
\psfrag{numax}{$\nu_{\rm max}$}
\psfrag{ar}{\small{$\ar$}}
\psfrag{sigma}{$\sigma$}
\psfrag{alpha_sigma}{$\alpha(\sigma)$}
\psfrag{time}{\scriptsize{$t$}}
\psfrag{time2}{\tiny{$t$}}
\psfrag{Q1}{\tiny{$Q_1$}}
\psfrag{Q2}{\tiny{$Q_2$}}
\psfrag{Q3}{\tiny{$Q_3$}}
\psfrag{Q4}{\tiny{$Q_4$}}
\psfrag{Q5}{\tiny{$Q_5$}}
\psfrag{b1}{\small{$\beta = 1$}}
\psfrag{b2}{\small{$\beta = 2$}}
\psfrag{b4}{\small{$\beta = 4$}}
\psfrag{b5}{\small{$\beta = 5$}}
\psfrag{b9}{\small{$\beta = 9$}}

\graphicspath{{Images/}}

\begin{document}
\title{Spatial fairness in linear wireless multi-access networks}
\author{P.M. van de Ven$^{1,2}$\\ D. Denteneer$^{3}$ \and J.S.H. van Leeuwaarden$^{1,2}$\\ A.J.E.M. Janssen$^{2,4}$ }

\footnotetext[1]{Eindhoven University of Technology, Department of Mathematics and Computer Science, P.O. Box 513, 5600 MB Eindhoven, The Netherlands}
\footnotetext[2]{Eurandom, P.O. Box 513, 5600 MB Eindhoven, The Netherlands}
\footnotetext[3]{Philips Research Europe, 5656 AA Eindhoven, The Netherlands}
\footnotetext[4]{Eindhoven University of Technology, Department of Electrical Engineering, P.O. Box 513, 5600 MB Eindhoven, The Netherlands}

\maketitle

\begin{abstract}
Multi-access networks may exhibit severe unfairness in throughput. Recent studies show that this unfairness is due to local differences in the neighborhood structure: Nodes with less neighbors receive better access. We study the unfairness in saturated linear networks, and adapt the multi-access CSMA protocol to remove the unfairness completely, by choosing the activation rates of nodes appropriately as a function of the number of neighbors. We then investigate the consequences of this choice of activation rates on the network-average saturated throughput, and we show that these rates perform well in a non-saturated setting.
\end{abstract}


%
%

\section{Introduction}
Multi-access protocols such as CSMA~\cite{KlTo75} as used in the IEEE 802.11 standard have gained much popularity for their ability to regulate the access of network nodes to a shared communication channel in a simple and fully distributed fashion. A major drawback of these protocols, however, is that they can exhibit severe unfairness, in the sense that some of the nodes get starved, while others receive good access. One of the main causes of this unfairness is that all nodes are treated the same, irrespective of their position in the network. We propose a way of compensating for the possible spatial disadvantages of nodes by enhancing the multi-access protocols with local information about the immediate neighborhood of nodes.

Unfairness in wireless networks is an active topic of research.
Wang and Kar~\cite{WaKa05} considered three nodes on a line that only block their direct neighbors, and showed that the middle node is starved when the activation rate of all three nodes increases. Such unfairness has been studied for more general networks by Durvy {\em et al.}~\cite{DuDoTh07,DuDoTh09} and Denteneer {\em et al.}~\cite{DeBoVeHi08}. We study the same model as in \cite{DuDoTh07,DuDoTh09,DeBoVeHi08}, a network with $n$ nodes on a line, in which active nodes block a certain subset of other nodes. Unblocked nodes become active, and active nodes deactivate, after exponential times. Under these assumptions, the $n$-dimensional process that describes the activity of nodes is a reversible continuous-time Markov process that possesses an elegant product-form expression for its stationary distribution. So this idealized linear network strikes a good balance between tractability and complexity, as it retains the essential features of competition and unfairness.

We assume that an active node blocks the first $\beta$~nodes on both sides, and we say that nodes that might block each other are {\em neighbors}. Results from~\cite{DuDoTh07,DuDoTh09,DeBoVeHi08} suggest that the unfairness observed in this model is due entirely to boundary effects. To deal with the unfairness, we shall modify the model in one important way: Instead of letting nodes activate at the same rate, we allow for node-specific activation rates. By choosing the activation rate $\load_i$ of node $i$ as a particular function of the number of its neighbors, we can guarantee that all nodes in the network have the same throughput, completely removing the unfairness. Our main contribution is that we prove that this fair choice of activation rates $\load_i=\fsolv_i$ takes on the extremely simple form
\begin{equation}\label{eqn:fair_load}
\fsolv_i = \alpha (1 + \alpha)^{\gamma(i) - \gamma(1)}
\end{equation}
with $\gamma(i)$ the number of neighbors of node $i$, and $\alpha$ any positive constant. We note that these node-dependent activation rates are still in line with the distributed nature of the multi-access protocol, as $\fsolv_i$ only requires the number of neighbors, which a node can obtain locally by sensing its direct environment. By choosing the activation rates according to~\eqref{eqn:fair_load} we essentially ensure long-term fairness. The first result in this direction is due to Kelly~\cite{Kelly85}, who considered a tree network (of which the linear network is a special case) with nearest-neighbor blocking ($\beta = 1$).

If we choose the activation rates as in~\eqref{eqn:fair_load} and let $\alpha \rightarrow \infty$, the throughput approaches $1/(\beta + 1)$, the highest possible per-node throughput. Recently, Jiang and Walrand~\cite{JiWa08}, Rajagopalan and Shah~\cite{RaSh08} and Liu {\em et al.}~\cite{LiYiPrChPo09} proposed protocols for determining the activation rates that can attain this maximal throughput in general networks. To this end, they introduced adaptive mechanisms that converge over time to the throughput-optimal activation rates. The main difference between our approach and the results in~\cite{JiWa08,RaSh08,LiYiPrChPo09} is that we obtain an explicit expression for the throughput-optimal and fair activation rates, rather than relying on an algorithm to find these. In fact, our explicit solution can be used as a benchmark for validating, in this particular case, the various algorithms proposed in~\cite{JiWa08,RaSh08,LiYiPrChPo09}. Besides giving insight into the operation of the network, these explicit rates also provide immediate optimal performance of the network, rather than first going through a startup stage during which nodes have to learn the right activation rates. The results discussed in our paper are only valid for linear networks, although the Markov random field approach from~\cite{Kelly85} allows us the extend our results to various other topologies, see Section~\ref{sec:conclusions}.

The paper is structured as follows. In Section~\ref{sec:model} we introduce the linear network in more detail. In Section~\ref{sec:unfairness} we study some of the key features of the unfairness that arises when all nodes have equal activation rates. Section~\ref{sec:fairness} contains the proof of the proposition that the activation rates in \eqref{eqn:fair_load} yield equal throughputs. In Section~\ref{sec:throughput} we investigate the impact of the rates in~\eqref{eqn:fair_load} on the network-average throughput, and in Section~\ref{sec:non-saturated} we discuss the extension to non-saturated networks. Section~\ref{sec:conclusions} presents some conclusions and further research directions.

\section{Model description}\label{sec:model}

We consider $n$ nodes on a line, and assume that all nodes are saturated (i.e.\ have an infinite supply of packets available for transmission). Nodes activate according to a $\beta$-hop blocking model, so a transmitting node blocks the first $\beta$~nodes on both sides. When node~$i$ is blocked, it remains silent until all nodes within distance~$\beta$ are inactive, at which point it tries to activate after an exponentially distributed (backoff) time with mean $1/\load_i$. Node~$i$ activates if it is still unblocked when the backoff timer runs out. If a node finds itself blocked when the backoff timer expires, it waits until all neighboring nodes are inactive once more and then repeats the backoff procedure. Without loss of generality, we assume that transmissions last for an exponentially distributed time with unit mean. Under these assumptions, the $n$-dimensional process that describes the activity of nodes is a continuous-time Markov process.

Each state of the Markov process is described as
$$
\omega = (\omega_1,\dots,\omega_n) \in \{0,1\}^n,
$$
where $\omega_i = 1$ when node~$i$ is active. Let $\Omega \subseteq \{0,1\}^n$ be the set of all feasible states.
Call $\omega\in\Omega$ {\it feasible} if no two $1$'s in $\omega$ are $\beta$ positions or less apart, i.e.,~$\omega_i\omega_k=0$ if  $1 \le |i-k|\le \beta$.

The Markov process that describes the activity of nodes is then fully specified by the state space $\Omega$ and the transition rates
\begin{equation}
r(\omega,\omega')=\left\{
                    \begin{array}{ll}
                      \load_i & \hbox{if $\omega'=\omega + e_i$,} \\
                      1 & \hbox{if $\omega'=\omega - e_i$,} \\
                      0 & \hbox{otherwise}.
                    \end{array}
                  \right.
\end{equation}
Here $e_i$ denotes the vector with all zeros except for a 1 at position $i$.

Alternatively, we can express the set of feasible states as all states that satisfy a certain system of linear equations. Let $A$ be an $(n-\beta) \times n$ matrix where each row contains $\beta+1$ consecutive 1's, in the following way:


\begin{equation}\label{eqn:Matrix_A}
A = \left(
\begin{array}{cccccccc}
1   &    1  &   \dots   &   1   &   0  &   \dots   &  0 &   0           \\
0   &   1  &   1    &   \dots   &   1   &   0   &   \dots               &   0           \\
    &    \multicolumn{3}{c}{\ddots} &      \multicolumn{3}{c}{\ddots}   &   \vdots   \\
0   &   \dots   &     0 &  1     &   1   &  \dots   &   1   &      0\\
0   &   0   &   \dots   &     0 &  1    &   1  &   \dots  &   1
\end{array}
\right).\end{equation}

Now we can write the state space as $\Omega = \{\omega \in \{0,1\}^n \mid A \omega \leq C\}$, where $C$ is the all-1 vector of size~$n$. This characterization has a natural interpretation as a set of capacity constraints, and nodes can activate only when enough capacity is available. We allocate unit capacity to each node, and use the convention that whenever a node is active it uses its own capacity, as well as the capacity of all its neighbors to the left. The $i$-th row of $A$ thus represents the capacity required when node~$i$ is active. The constraints that arise from the last $\beta$ nodes on the line are redundant, and ignoring these leads to the matrix $A$ in~\eqref{eqn:Matrix_A}.

From the description of the state space as a set of capacity constraints, it is clear that our model belongs to the general class of loss networks, see Kelly~\cite{Kelly91}. Loss networks are known to be reversible, and possess product-form solutions. For our Markov process, this product-form solution is given by the stationary measure $\pi$ on $\Omega$ for which
\begin{equation}\label{eqn:lim_dist}
\pi(\omega)=\left\{
         \begin{array}{ll}
           Z_n^{-1}\prod_{i=1}^n\load_i^{\omega_i} & \hbox{if $\omega$ is feasible,} \\
           0 & \hbox{otherwise,}
         \end{array}
       \right.
\end{equation}
and where $Z_n$ is the normalization constant that makes $\pi$ a probability measure. This result is well-known in the context of wireless networks, see e.g.~\cite{BoKe80,DeBoVeHi08,DuDoTh07,WaKa05}.

Our main concern is with the long-term behavior of nodes, characterized by their throughputs. A common throughput-degrading phenomenon in wireless networks is collisions, which may occur when multiple nearby nodes transmit simultaneously, causing these transmissions to fail. However, we assume that the blocking range~$\beta$ is large enough to rule out collisions, so any activity of a node contributes to its throughput. Although a model without collisions might seem limited, numerous simulation studies show that choosing the blocking range just large enough to preclude collisions gives very good performance, see e.g.~\cite{HiZaMaJuWeDeBeMa08,VaRaWo05,YaVa05,ZhGuYaCo04}. In fact, the forthcoming paper~\cite{JaLeVe09} shows that this choice is throughput-optimal when the activation rates are sufficiently large. We study the throughput vector $\theta = (\theta_1,\dots,\theta_n)$, where $\theta_i$ represents the fraction of time node $i$ is active. Denote the total number of feasible states by $K$, let $\Omega=\{\Omega_1,\ldots,\Omega_K\}$, and introduce the $n \times K$ incidence matrix $X$ such that $X_{ik} = 1$ when the $i$th element in the state $\Omega_k$ equals 1. Then
\begin{equation}\label{eqn:throughput}
\theta = X \cdot \Pi,
\end{equation}
with $\Pi=(\pi(\Omega_1),\ldots,\pi(\Omega_K))$.

By exploiting the structure of the network, we can construct alternative expressions for the throughput in~\eqref{eqn:throughput}. Specifically, we shall make use of the observation that if node~$i$ is active, nodes to the left of~$i$ behave independently from nodes to the right of~$i$. This leads to the following theorem.
\begin{theorem}\label{thm:throughput}
Define the sequence $(Z_i)_{i = -\infty}^\infty$ such that $Z_i = 1$ for $i \le 0$, and
\begin{align}
\label{eqn:def_Z_small} Z_i &= 1 + \load_1 + \dots + \load_i, \quad i = 1,2,\dots,\beta + 1,\\
\label{eqn:def_Z_large} Z_i &= Z_{i-1} + \load_i Z_{i-\beta-1}, \quad i =  \beta + 2,\beta + 3,\dots.
\end{align}
Let the vector of activation rates $\load = (\load_1,\dots,\load_n)$ be symmetric. Then
\begin{equation}\label{eqn:throughput_alternative}
\theta_i = \load_i \frac{Z_{i-\beta-1} Z_{n-i-\beta}}{Z_n}, \quad i=1,\dots,n.
\end{equation}
\end{theorem}

\begin{proof}
By conditioning on whether node~$i$ is active, we can decompose the activity of the network into two parts, separated by this active node (see~\cite{BoKe80}, Equation~(15)),
\begin{equation}\label{eqn:throughput_Boorstyn}
\theta_i = \load_i \frac{Z_{1:i-\beta-1} Z_{i +\beta + 1:n}}{Z_{1:n}},
\end{equation}
where $Z_{i:j}$ is the normalization constant of a network consisting only of nodes $i,\dots,j$. For simplicity we denote $Z_i:=Z_{1:i}$, and the symmetry of $\load$ implies
\begin{equation}\label{eqn:Z_i_symmetric}
Z_{i:n} = Z_{1:n-i+1}.
\end{equation}
Substituting~\eqref{eqn:Z_i_symmetric} into~\eqref{eqn:throughput_Boorstyn} yields the expression for $\theta_i$ in~\eqref{eqn:throughput_alternative}. By conditioning on the activity of node~$i$, we immediately get the recursion relation~\eqref{eqn:def_Z_large} for the $Z_i$.
\end{proof}


%
%
%
%
%
%
%
%

\section{Unfairness}\label{sec:unfairness}

We now venture deeper into the issue of unfairness, and assume all nodes to have equal activation rates $\load_i = \sigma$. As observed by Durvy {\em et al.}~\cite{DuDoTh07} and Denteneer {\em et al.}~\cite{DeBoVeHi08}, the throughput distribution in this setting is highly unfair, in the sense that some nodes have a systematically higher throughput than others. This unfairness can be explained by the {\it node-in-the-middle} phenomenon discussed for example in Wang and Kar~\cite{WaKa05} and Garetto {\em et al.}~\cite{GaSaKn06} for the case $n=3$, $\beta=1$: The middle node is in an unfavorable position as it has to wait for both outer nodes to deactivate, whereas these boundary nodes only compete for transmission with the middle node.


\begin{figure}[h!!]
 \begin{center}
 \subfigure[$n=6$]{\label{fig:unfair61} \includegraphics[width = 0.35\linewidth]{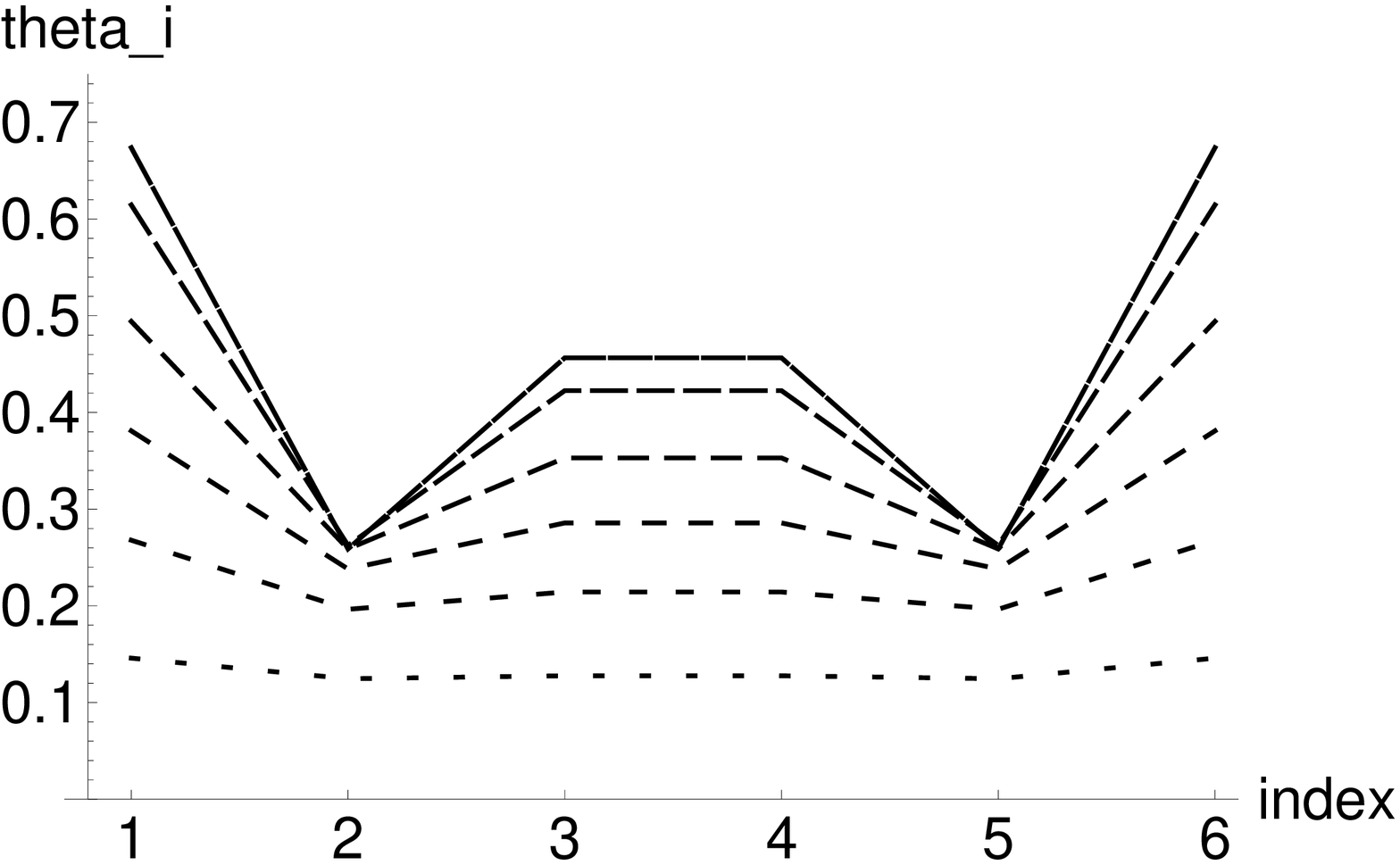}} \hspace{0.5cm}
 \subfigure[$n=9$]{\label{fig:unfair91} \includegraphics[width = 0.35\linewidth]{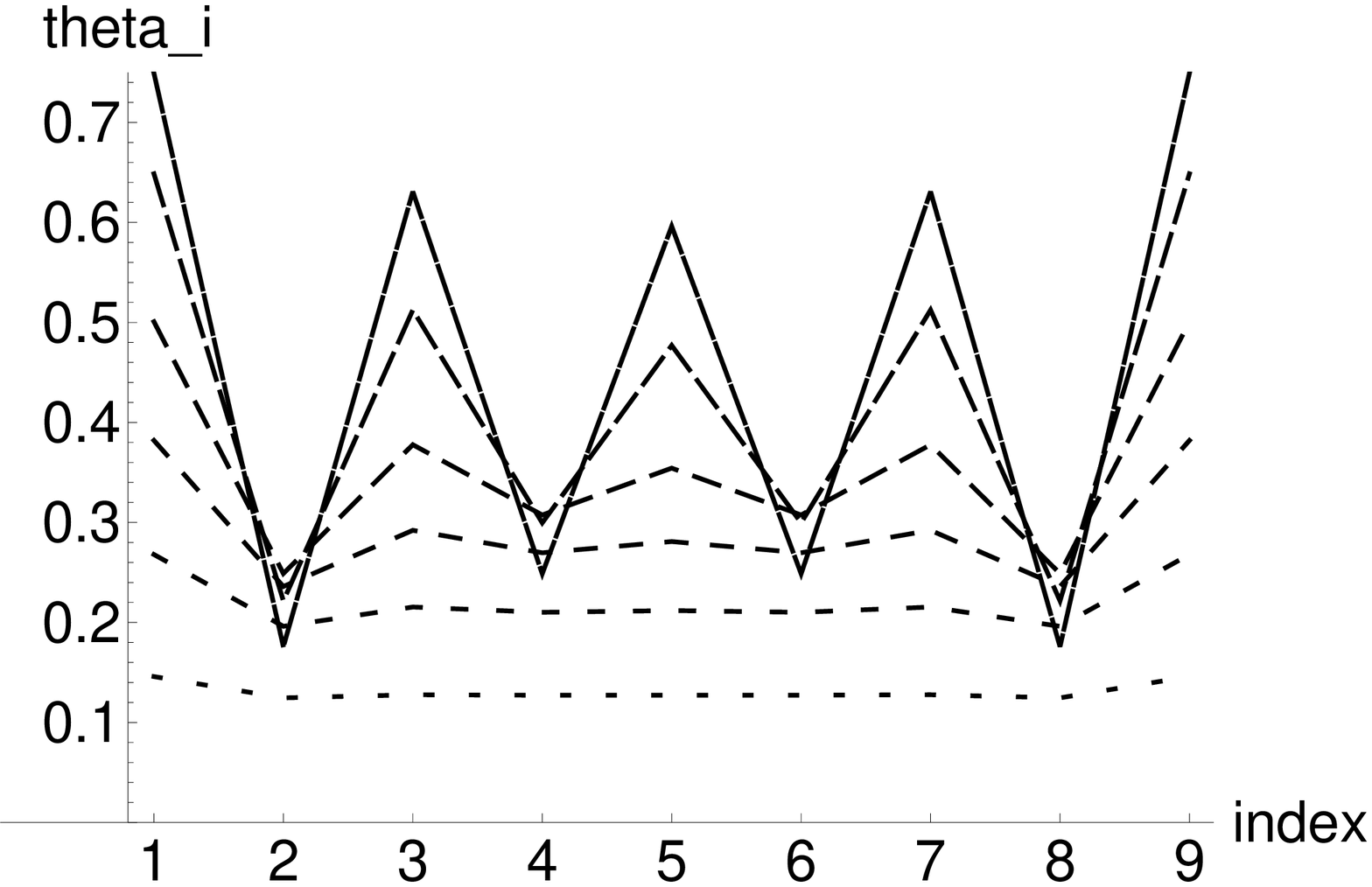}}
 \subfigure[$n=12$]{\label{fig:unfair121} \includegraphics[width = 0.35\linewidth]{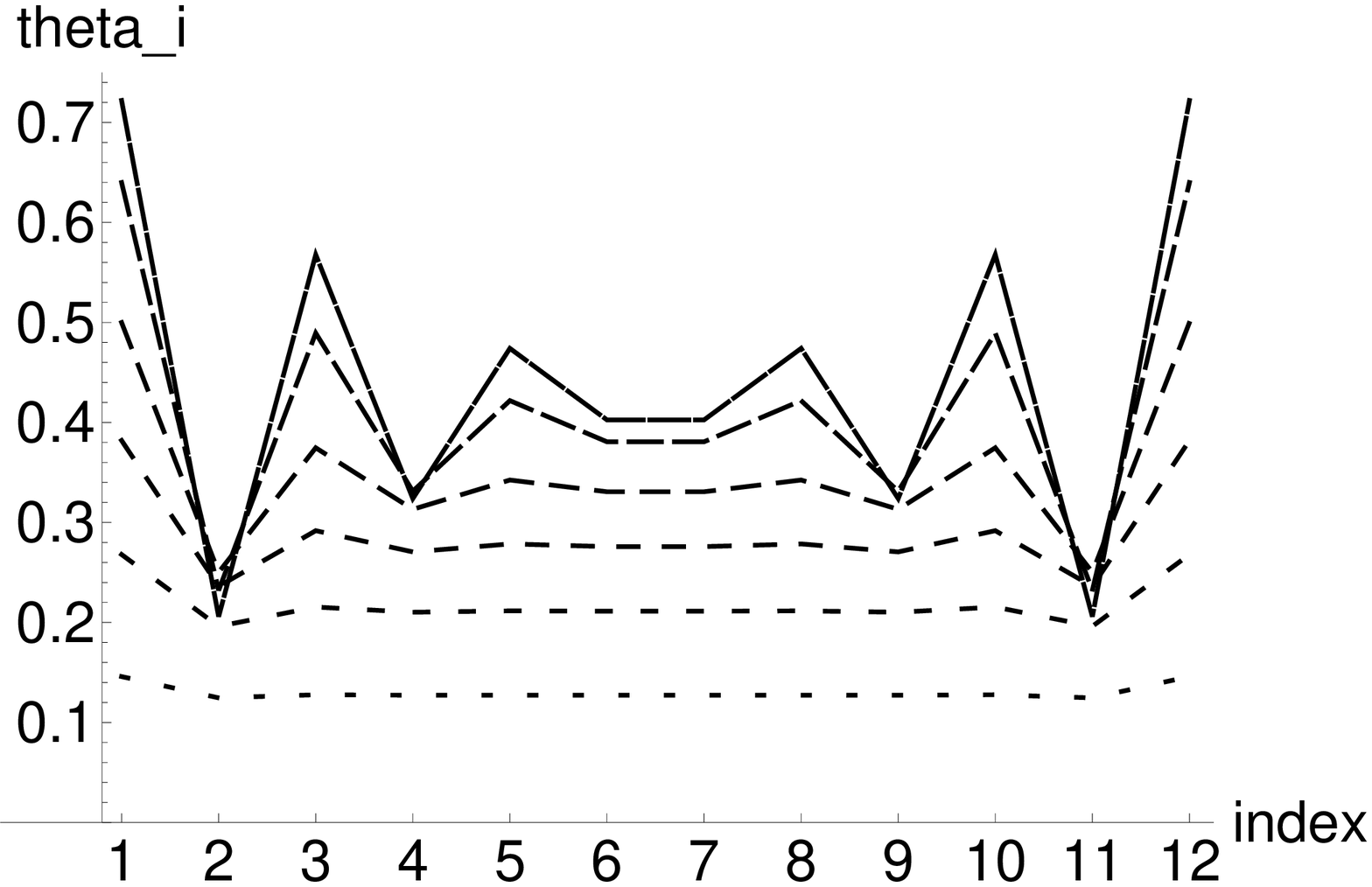}}\hspace{0.5cm}
 \subfigure[$n=15$]{\label{fig:unfair151} \includegraphics[width = 0.35\linewidth]{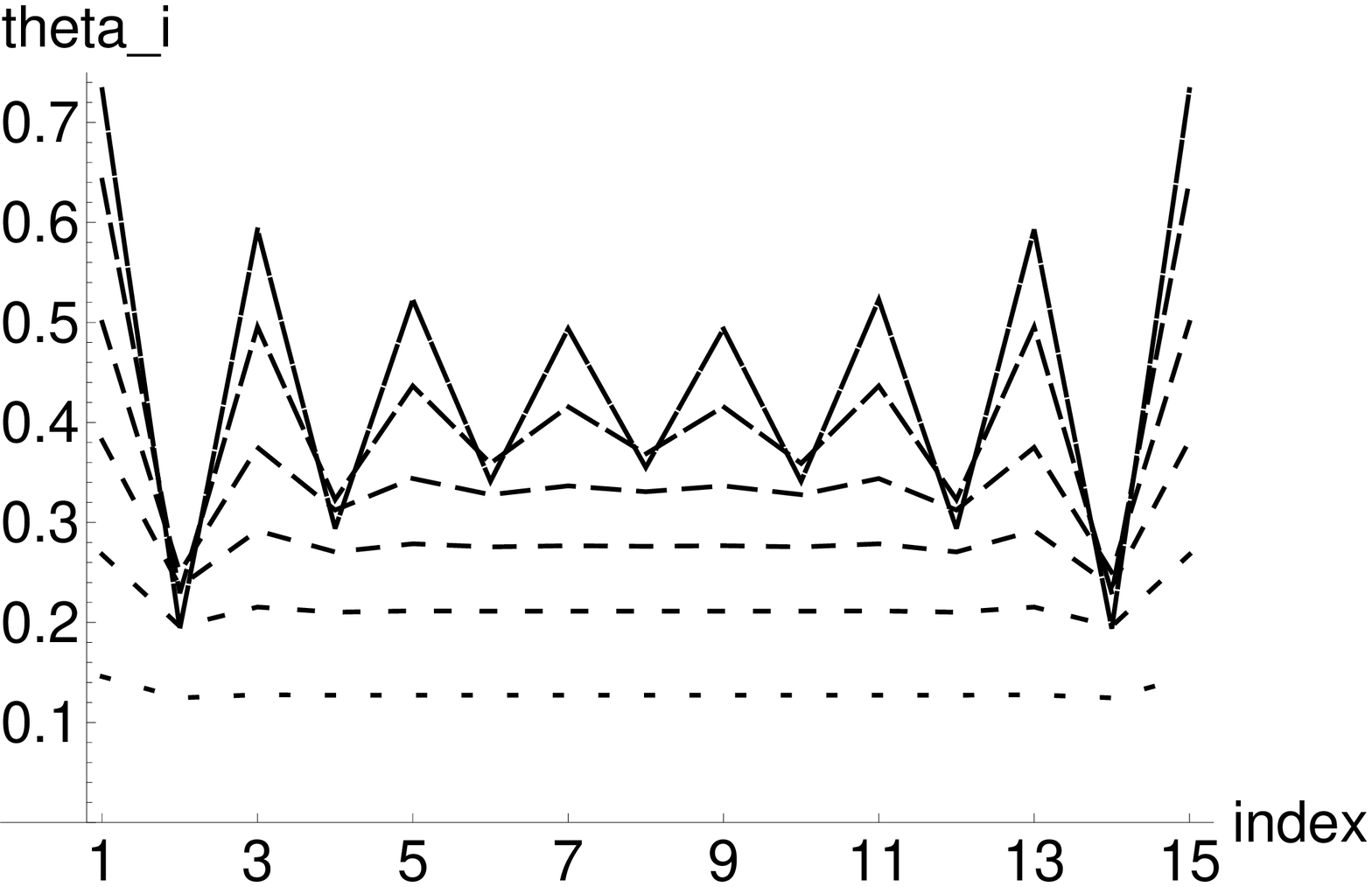}}
 \subfigure{ \includegraphics[width = 0.4\linewidth]{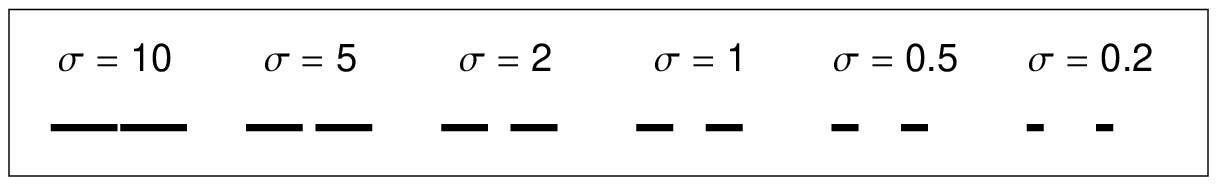}}
 \end{center}
  \caption{The per-node throughput for $\beta = 1$ and various values of $n$ and $\sigma$}
 \label{fig:unfair}
\end{figure}

In order to compute the throughput from~\eqref{eqn:throughput_alternative}, we require information on the $Z_i$. For general activation rates $\load_i$ these cannot be solved explicitly from the recursion~\eqref{eqn:def_Z_small}-\eqref{eqn:def_Z_large} (an exception to this are the rates in~\eqref{eqn:fair_load}, see Section~\ref{sec:fairness}). However, when all nodes have equal activation rate $\sigma$ we can rewrite the $Z_i$ as (see~\ref{sec:partition_function})
\begin{equation}\label{dfg2}
Z_i = \sum_{j = 0}^\beta c_j \lambda_j^i, \quad i =0,1,\dots,
\end{equation}
where $\lambda_j$ are the roots of
\begin{equation}\label{eqn:equation_lambda2}
\lambda^{\beta+1} - \lambda^\beta - \sigma = 0,
\end{equation}
and the $c_j$ are given by
\begin{equation}\label{eqn:c_j2}
c_j = \frac{\lambda_j^{\beta + 1}}{(\beta+1)\lambda_j - \beta}.
\end{equation}
Moreover, when the index grows large ($i \rightarrow \infty$), the normalization constant~\eqref{dfg2} is dominated by the largest root $\lambda_0$, i.e.,
\begin{equation}\label{eqn:partition_function_large2}
Z_i \sim c_0 \lambda_0^i, \quad i \rightarrow \infty.
\end{equation}
We shall use this asymptotic relation at several places, see~\eqref{eqn::diff12} and~\eqref{eqn:throughput_unfair_large_n_2}.

For ease of presentation, we restrict ourselves to $\beta=1$ for the remainder of this section. Figures~\ref{fig:unfair61}-\ref{fig:unfair151} show the per-node throughput for various values of $n$ and $\sigma$. All figures display a similar pattern, with the outer nodes having the highest throughput. Moreover, the throughput is symmetric in all figures, and exhibits some form of oscillatory behavior. These observations are formalized in the following result.
\begin{proposition}\label{pro:single_k=1}
For $\load_i = \sigma >0$, $i = 1,\ldots,n$ and $\beta = 1$, the throughput has the following properties:\\
(i) Symmetric: $\theta_i = \theta_{n - i + 1}$, $i = 1,2,\dots,n$.\\
(ii) Alternating and converging: $(-1)^i(\theta_{i+1} - \theta_i)$ is positive and decreasing for $i = 1,2,\ldots,\lfloor n/2 \rfloor$.
\end{proposition}
Proposition~\ref{pro:single_k=1} is proven in~\ref{app1}.

In Figure~\ref{fig:unfair} we see that for nearest-neighbor blocking, the largest difference in throughput is between $\theta_1$ and $\theta_2$. Proposition \ref{pro:single_k=1}(ii) confirms that this is the most unfair situation, and it persists even in larger networks where the node-in-the-middle problem is mitigated by the activity of the remaining nodes. In fact, for large networks we have the following result.

\begin{proposition}\label{pro:diff12} For $\load_i = \sigma >0$,\ $i = 1,\dots,n$ and $\beta = 1$,
\begin{equation}
\frac{\theta_1}{\theta_2}\sim\frac{1+\sqrt{1+4\sigma}}{2}, \quad n\rightarrow\infty.
\end{equation}
\end{proposition}

\begin{proof} We have that $\theta_1\propto Z_{n-2}$ and $\theta_2\propto Z_{n-3}$. Using~\eqref{eqn:partition_function_large2} we obtain
\begin{equation}\label{eqn::diff12}
\frac{\theta_1}{\theta_2} \sim \lambda_0, \quad n \rightarrow \infty.
\end{equation}
For $\beta = 1$ we can explicitly solve~\eqref{eqn:equation_lambda2} to obtain $\lambda_0$, and the result follows.
\end{proof}


We note that for $\beta = 1$, alternative descriptions of $Z_i$ exist of the forms
%
\begin{equation}
Z_i = (-\sigma)^{\frac{1}{2}(i+1)} U_{i+1}(\sqrt{-1/4 \sigma}),
\end{equation}
where $U_n(x)$ is the $n$th Chebyshev polynomial of the second kind, and
\begin{equation}
Z_i = \sum_{j = 0}^{\lfloor \frac{i+1}{2}\rfloor} {i+1-j \choose j} \sigma^j.
\end{equation}
The latter expression can be interpreted as the summation over all possible combinations of nodes that can be active simultaneously.

Figure~\ref{fig:unfair} shows another interesting property of this network. Increasing $\sigma$ leads in many cases to a higher throughput for each of the nodes. Hence, in such situations, one may want to increase $\sigma$ further. However, we also observe that there exists a critical value $\sigma^*$, such that at least one of the throughputs $\theta_i$ decreases as $\sigma$ increases beyond $\sigma^*$. The characterization of this critical value is a possible topic for future research.

\begin{figure}[h!]
 \begin{center}
 \subfigure[$\beta=2$]{\label{fig:unfair92} \includegraphics[width= 0.33\linewidth]{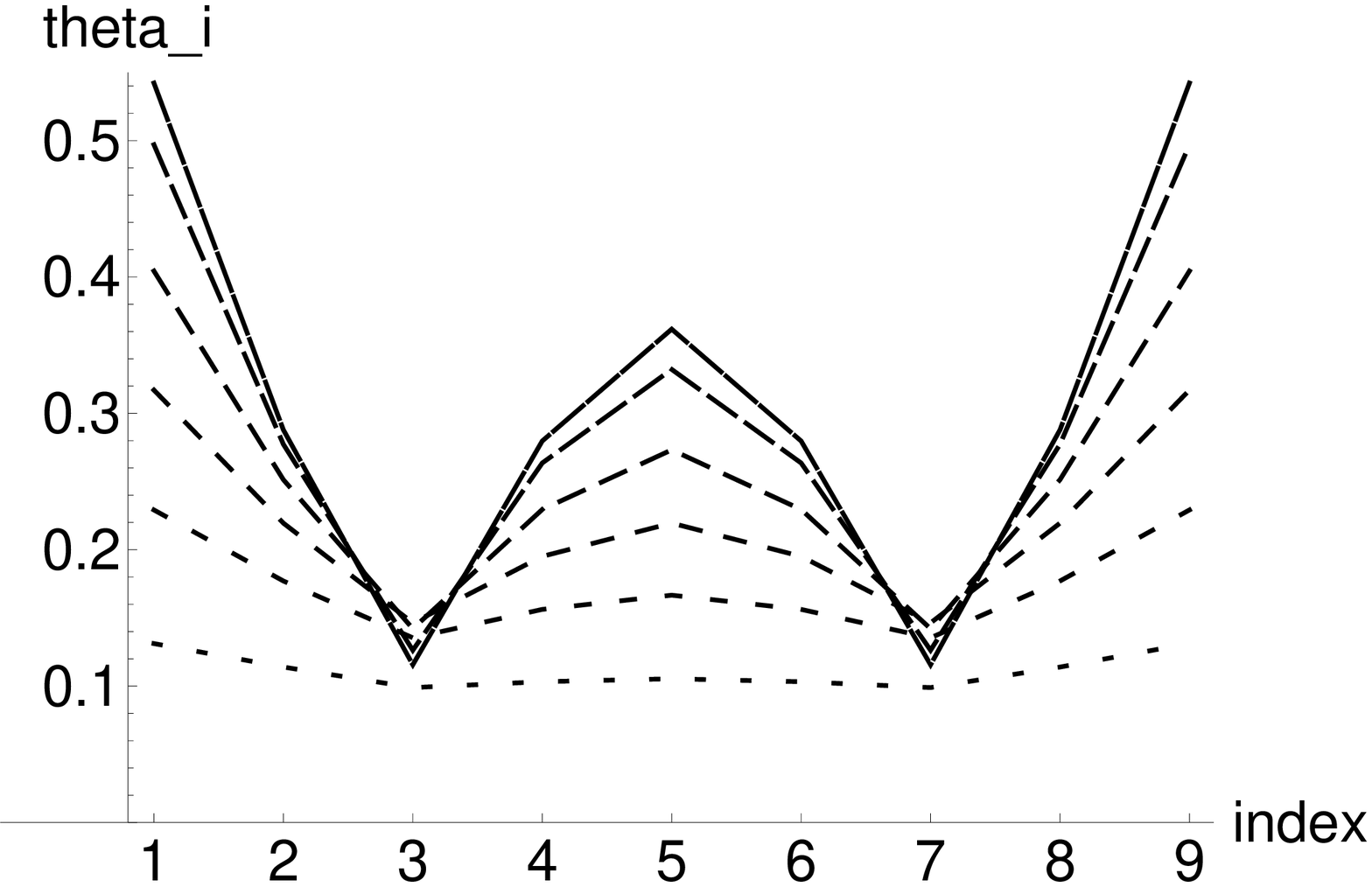}}\hspace{0.5cm}
 \subfigure[$\beta=3$]{\label{fig:unfair93} \includegraphics[width= 0.33\linewidth]{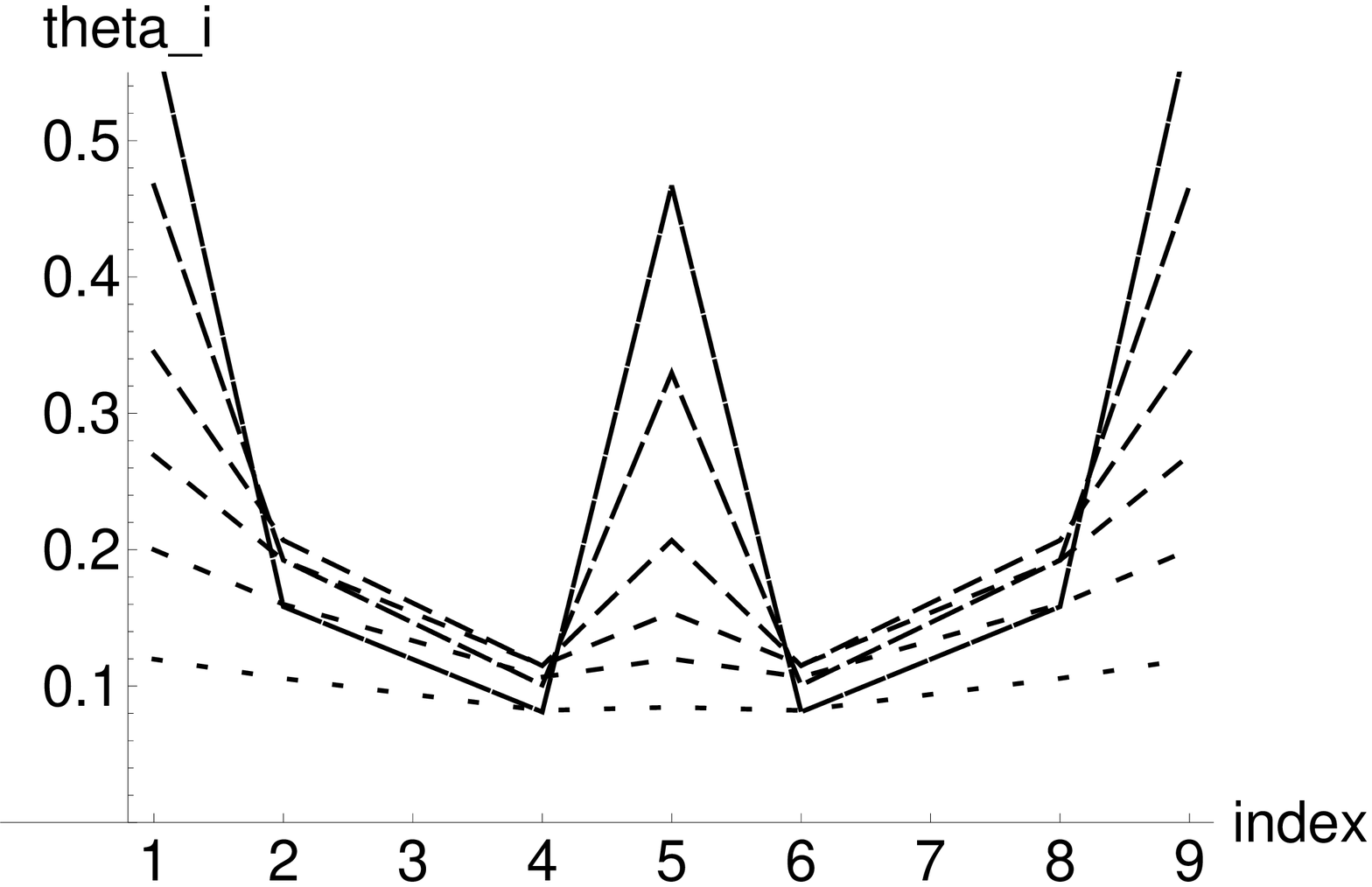}}
 \subfigure{\includegraphics[width = 0.4\linewidth]{legend}}
 \end{center}
  \caption{The per-node throughput for $n = 9$ and various values of $\beta$ and $\sigma$}
 \label{fig:unfair_kappa>1}
\end{figure}

Results similar to those presented in this section can be obtained for $\beta \ge 2$. As an example, Figures~\ref{fig:unfair92}-\ref{fig:unfair93} show the per-node throughput for $n=9$ and $\beta = 2,3$, obtained by iterating the recursion~\eqref{eqn:def_Z_small}-\eqref{eqn:def_Z_large}. Both figures exhibit similar oscillatory behavior as observed for $\beta = 1$, although the oscillation period increases with $\beta$.

\section{Fairness}\label{sec:fairness}
In this section we present a way to completely remove the unfairness that was addressed in Section~\ref{sec:unfairness}. In order to do so, we choose node-dependent activation rates $\load_i$ such that all nodes have equal throughput ($\theta_1 = \theta_2 = \dots = \theta_n$). From~\eqref{eqn:lim_dist} and~\eqref{eqn:throughput} we see that in order to meet this objective we have to solve a system of $n$ nonlinear equations. It seems that in general this system cannot be solved directly. We therefore choose a more indirect approach, for which we first consider two special cases that can be solved explicitly. The insight obtained from these exact solutions is then used to guess the general solution to the system of non-linear equations.

The first case is where $\beta=n-2$, so that all but the two outer nodes will block the entire network.
\begin{proposition}\label{pro:fair_Dee}
For linear networks with 3 or more nodes, and $\beta = n-2$, setting $\load_1 = \load_n = \alpha$ and $\load_i = \alpha(1 + \alpha)$ for all other nodes yields equal throughputs
\begin{equation}\label{eqn:fair_dee_throughput}
\theta_i = \frac{\alpha}{1 + (n-1) \alpha}, \quad i = 1,\dots,n.
\end{equation}
\end{proposition}
\begin{proof}
The expression for the throughput in~\eqref{eqn:throughput} can be written as
\begin{align}
\theta_1 &= Z_n^{-1} \load_1(1 + \load_n),\label{eqn:throughput_n_k=n-2_1}\\
\theta_i &= Z_n^{-1} \load_i, \quad i = 2,3,\dots,n-1, \label{eqn:throughput_n_k=n-2_2}\\
\theta_n &= Z_n^{-1} \load_n(1 + \load_1).
 \label{eqn:throughput_n_k=n-2_3}
\end{align}
The inherent symmetry of the model allows us to set $\load_1 = \load_n$. Moreover, for the throughput of the other nodes to be equal, we require $\load_2 = \dots = \load_{n-1} = \load_1 (1 + \load_1)$. If we set $\load_1 = \alpha$, and substitute this into~\eqref{eqn:throughput_n_k=n-2_1}-\eqref{eqn:throughput_n_k=n-2_3}, we get a throughput of
\begin{equation}\label{eqn:fair_dee_1}
\theta_i = Z_n^{-1} \alpha (1 + \alpha). 
\end{equation}
The normalization constant $Z_n$ can be determined by summing over all feasible states:
\begin{align}
\nonumber Z_n &= 1 + \sum_{i = 1}^n \load_i + \load_1 \load_n = 1 + (n-2) \alpha(1 + \alpha) + 2 \alpha  + \alpha^2\\
&= (1+ \alpha)(1 + (n-1) \alpha ). \label{eqn:fair_dee_2}
\end{align}
Substituting~\eqref{eqn:fair_dee_2} into~\eqref{eqn:fair_dee_1} yields~\eqref{eqn:fair_dee_throughput}.
\end{proof}

The case $n = 5$, $\beta = 3$  of Proposition~\ref{pro:fair_Dee} was considered in \cite{DeBoVeHi08}. The second special case corresponds to $n = 2(\beta + 1)$, so that a node blocks at least half of the network.
\begin{proposition}\label{pro:fair_Johan}
For linear networks with $n = 2m$ nodes, $m \in \mathds{N}$, and $\beta = m-1$, setting $\load_i = \alpha(1 + \alpha)^{i-1}$  for $i = 1,\dots,m$ yields equal throughputs
\begin{equation}\label{eqn:fair_Johan_throughput}
\theta_i = \frac{\alpha}{1 + m \alpha}, \quad i = 1,\dots,n.
\end{equation}
\end{proposition}

\begin{proof}
To achieve equal throughputs, we see from~\eqref{eqn:lim_dist} and~\eqref{eqn:throughput} that for the case at hand we should solve the system of equations
\begin{align}
\nonumber \load_1 + \load_1 (\load_{m+1} + \dots + \load_n) &= \load_2 + \load_2(\load_{m+2} + \dots + \load_n)\\
\nonumber &= \load_3 + \load_3(\load_{m+3} + \dots + \load_n)\\
\nonumber &\hspace{0.5em}\vdots\\
&= \load_m + \load_m \load_n. \label{eqn:fair_Johan_1}
\end{align}
Indeed, the throughput of node~$i$ can be written as a sum over all states in which node~$i$ is active. Using symmetry,~\eqref{eqn:fair_Johan_1} can be written as
\begin{align}
\nonumber \load_1 + \load_1 (\load_{1} + \dots + \load_m) &= \load_2 + \load_2(\load_{1} + \dots + \load_{m-1})\\
\nonumber &= \load_3 + \load_3(\load_{1} + \dots + \load_{m-2})\\
\nonumber &\hspace{0.5em}\vdots\\
&= \load_m + \load_m \load_1. \label{eqn:fair_Johan_2}
\end{align}
Let $\load_1 = \alpha >0$. The solution of~\eqref{eqn:fair_Johan_2} is easily seen to be
\begin{equation}
\load_i = \alpha(1 + \alpha)^{i-1},\quad i = 1,\dots,m,
\end{equation}
and hence
\begin{equation}\label{eqn:fair_Johan_3}
\quad \theta_i = Z_n^{-1} \alpha(1 + \alpha)^m.
\end{equation}
Summing over all possible states yields
\begin{align}
\nonumber Z_n &= 1 + \sum_{i = 1}^n \load_i + \sum_{i = 1}^m \load_i \sum_{j = i+m}^n \load_j\\
\nonumber &= 1 + \sum_{i = 1}^m \load_i + \sum_{i = 1}^m \load_i (1 + \load_1 + \dots + \load_{m-i-1})\\
\nonumber &= 1 + ((1 + \alpha)^m - 1) + m \alpha (1 + \alpha)^m\\
\label{eqn:fair_Johan_4} &= (1 + m \alpha)(1 + \alpha)^m.
\end{align}
Substituting~\eqref{eqn:fair_Johan_4} into \eqref{eqn:fair_Johan_3} gives~\eqref{eqn:fair_Johan_throughput}.
\end{proof}


It is clear that the complexity of the system of equations governed by~\eqref{eqn:throughput} reduces considerably for the choices of $\beta$ discussed in Propositions \ref{pro:fair_Dee} and~\ref{pro:fair_Johan}. For general $\beta$ this system remains rather complicated, and the direct approach taken in the proofs of Propositions \ref{pro:fair_Dee} and~\ref{pro:fair_Johan} no longer seems to work.

However, we can use Propositions~\ref{pro:fair_Dee} and~\ref{pro:fair_Johan} to make an educated guess about the general solution. First observe that the fair activation rates in Propositions~\ref{pro:fair_Dee} and~\ref{pro:fair_Johan} only depend on the number of neighbors (nodes within $\beta$ hops) that each node has. Denote by $\gamma(i)$ the number of neighbors of node $i$, let $\alpha > 0$, and choose activation rates $\fsolv_i$ as in~\eqref{eqn:fair_load}. We see that this choice is consistent with the fair activation rates in Propositions~\ref{pro:fair_Dee} and~\ref{pro:fair_Johan}. We now show that $\fsolv_i$ indeed achieves fairness for all $\beta$. To this end, we first show that when the activation rates are chosen according to~\eqref{eqn:fair_load}, the recursive relation~\eqref{eqn:def_Z_small}-\eqref{eqn:def_Z_large} for the normalization constant $Z_i$ has a closed-form solution.


\begin{lemma}\label{lem:H}
Let $\alpha > 0$ and choose $\fsolv_i$ as in~\eqref{eqn:fair_load}. Then
\begin{equation}\label{eqn:norm_const}
Z_i = (1 + \alpha)^i,~~~i  = 1,2,\dots, n  - \beta.
\end{equation}
\end{lemma}

\begin{proof}
This can be verified by substituting the solution~\eqref{eqn:norm_const} into the relations~\eqref{eqn:def_Z_small} and~\eqref{eqn:def_Z_large}. For $i \le \beta +1$,~\eqref{eqn:def_Z_small} gives
\begin{equation}
Z_i = 1 + \alpha + \alpha (1 + \alpha) + \dots + \alpha (1 + \alpha)^{i-1} = (1 + \alpha)^i.
\end{equation}
From~\eqref{eqn:def_Z_large} we see that
\begin{equation}
Z_i = (1 + \alpha)^{i-1} + \alpha (1 + \alpha)^\beta (1 + \alpha)^{i-\beta -1} = (1 + \alpha)^i,
\end{equation}
covering the case $i \ge \beta + 2$.
\end{proof}

With Lemma~\ref{lem:H} we are now in the position to prove our main result.
\begin{theorem}\label{thm:fair_allocation}
Let $\alpha > 0$, $\beta \le n-1$ and choose $\fsolv_i$ as in~\eqref{eqn:fair_load}. Then
\begin{equation}\label{eqn:throughput_fair}
\theta_i = \frac{\alpha}{1 + (1 + \beta) \alpha}, \quad i = 1,\dots,n.
\end{equation}
\end{theorem}

\begin{proof}
To prove this result we substitute the normalization constants from Lemma~\ref{lem:H} into the expression for the throughput in \eqref{eqn:throughput_alternative}. We distinguish between different values of $i$.

For $i \ge \beta + 1$ and $i \le n-\beta$ we see that $\fsolv_i = \alpha (1 + \alpha)^{\beta}$ and
\begin{equation}\label{eqn:proof_fair_allocation4}
Z_{i - \beta - 1} = (1 + \alpha)^{i - \beta - 1}, \quad Z_{n - i - \beta} = (1 + \alpha)^{n - i - \beta}.
\end{equation}
Similarly for $i \ge \beta + 1$ and $i \ge n-\beta + 1$ we have $\fsolv_i = \alpha (1 + \alpha)^{n - i}$ and
\begin{equation}\label{eqn:proof_fair_allocation5}
Z_{i - \beta - 1} = (1 + \alpha)^{i - \beta - 1}, \quad Z_{n - i - \beta} = 1.
\end{equation}
For $i \le \beta$ and $i \le n-\beta$ we have $\fsolv_i = \alpha (1 + \alpha)^{i-1},$ and
\begin{equation}\label{eqn:proof_fair_allocation6}
Z_{i - \beta - 1} = 1, \quad Z_{n - i - \beta} = (1 + \alpha)^{n - i - \beta},
\end{equation}
and finally for $i \le \beta$ and $i \ge n-\beta + 1$ we have $\fsolv_i = \alpha (1 + \alpha)^{n -\beta - 1}$ and
\begin{equation}\label{eqn:proof_fair_allocation7}
Z_{i - \beta - 1} = 1, \quad Z_{n - i - \beta} = 1.
\end{equation}
Substituting \eqref{eqn:proof_fair_allocation4}-\eqref{eqn:proof_fair_allocation7} into~\eqref{eqn:throughput_alternative} yields
\begin{equation}
\theta_i = Z_n^{-1} \alpha (1+\alpha)^{n-\beta-1}.
\label{eqn:proof_fair_allocation2}
\end{equation}

We next consider the normalization constant. Choose $m$ such that $n = \beta + m$. Then
\begin{align}
\nonumber Z_n ={}& Z_{n-1} + \fsolv_n Z_{n-\beta-1}\\
\nonumber ={}& Z_{n-2} + \fsolv_{n-1} Z_{n-\beta-2} + \fsolv_n Z_{n-\beta-1}\\
\nonumber \vdots{}\hspace{0.2cm}&\\
\label{eqn:proof_fair_allocation8} ={}& Z_{n-\beta} + \sum_{i = 1}^{\beta} \fsolv_{n+1-i} Z_{n-\beta-i}.
\end{align}
Substituting~\eqref{eqn:norm_const} into~\eqref{eqn:proof_fair_allocation8} yields
\begin{align}
\nonumber Z_n ={}& (1 + \alpha)^{n - \beta} + \sum_{i = 1}^{\min\{m,\beta\}} \alpha (1 + \alpha)^{i-1} (1+\alpha)^{n-\beta-i} + \sum_{i = m+1}^{\beta} \alpha (1 + \alpha)^{n - \beta - i}\\
\label{eqn:proof_fair_allocation3} ={}& (1 + \alpha)^{n - \beta - 1} (1 + (\beta+1) \alpha).
\end{align}
Combining \eqref{eqn:proof_fair_allocation3} and \eqref{eqn:proof_fair_allocation2} leads to~\eqref{eqn:throughput_fair}.
\end{proof}

To better understand why the rates~\eqref{eqn:fair_load} successfully ensure strict fairness using only the number of neighbors of each node, we study its behavior in the two limiting regimes of light traffic ($\alpha \downarrow 0$) and heavy traffic $(\alpha \rightarrow \infty)$. First rewrite~\eqref{eqn:fair_load} as
\begin{equation}\label{eqn:fair_load_alternative}
\fsolv_i = \alpha \sum_{j = 0}^{\gamma(i) - \gamma(1)} {\gamma(i) - \gamma(1) \choose j} \alpha^j, \quad i  = 1,\dots,n.
\end{equation}

When $\alpha$ is small, nodes activate slowly, and few nodes will be active simultaneously. In fact, the limiting distribution is dominated by states with at most one node active, and node interaction (blocking) is negligible. This is reflected in the light-traffic activation rates that follow immediately from~\eqref{eqn:fair_load_alternative}:
\begin{equation}\label{eqn:activity_light_traffic}
\fsolv_i = \alpha + (\gamma(i) - \gamma(1)) \alpha^2 + \mathcal{O}(\alpha^3),\quad \alpha \downarrow 0.
\end{equation}
So the first-order light-traffic activation rate $\fsolv_i = \alpha$ is the same for all nodes. Indeed, if at most one node is active (as is the case for $\alpha$ small), there is no blocking, and therefore no need to differentiate among nodes. As $\alpha$ increases, states with two active nodes are increasingly likely. This introduces node interaction, as nodes may now block their neighbors (all nodes within distance $\beta$). This is accounted for in the activation rate by the term $(\gamma(i) - \gamma(1)) \alpha^2$, which is linear in the number of neighbors. So in light-traffic, only the number of neighbors is of importance, rather than the structure of the entire network. This reasoning extends to more general networks.

When $\alpha$ is large, we see from~\eqref{eqn:fair_load_alternative} that
\begin{equation}\label{eqn:hevay_traffic_activation}
\fsolv_i = \alpha^{\gamma(i) - \gamma(1) + 1} + \mathcal{O}(\alpha^{\gamma(i) - \gamma(1)}),\quad \alpha \rightarrow \infty.
\end{equation}
Nodes activate almost immediately when they get the chance to do so (i.e.\ when all neighbors are inactive), and thus the system spends almost all the time in maximal independent sets of active nodes. However, as some nodes have a higher activation rate than others, the only states $\omega$ that have positive limiting probability for $\alpha \rightarrow \infty$ are those that maximize the sums of the exponents of $\alpha$ over all active nodes, $\sum_{i = 1}^n (\gamma(i) - \gamma(1))\indi{\omega_i = 1}$, with $\indi{}$ the indicator function. This is in contrast to the heavy-traffic behavior of the system in which all nodes have equal activation rates $\sigma$. Here only states with the most active nodes (maximal independent sets) have positive limiting probability when $\sigma \rightarrow \infty$. As~\eqref{eqn:fair_load} provides fairness even in heavy-traffic, the heavy-traffic fair activation rates are such that all nodes are contained an equal number of times in these dominant sets.
This result strongly depends on the structure of the network, as the maximal independent sets may change drastically with the addition or the removal of even a single node. So contrary to light-traffic, the simple, locally determined heavy-traffic activation rates~\eqref{eqn:hevay_traffic_activation} may not easily extend beyond linear networks.


\section{Network-average throughput}\label{sec:throughput}

The fair rates $\fsolv_i$ in~\eqref{eqn:fair_load} are designed to remove the unfairness that arises when all nodes have equal activation rates $\sigma$.
In order to compare the two schemes, we want to set their respective parameters $\alpha$ and $\sigma$ such that the average per-node throughputs are equal.
For given equal rates $\sigma > 0$ we write $Z_i(\sigma)$ and $\theta_i(\sigma)$ for the normalization constant of a network with $i$ nodes, and the throughput of node~$i$, respectively. We let $\avrth_n(\sigma) := n^{-1} \sum_{i = 1}^n \theta_i(\sigma)$ denote the average per-node throughput in a network with $n$ nodes, and we denote $\avrth(\sigma):=\lim_{n \rightarrow \infty} \avrth_n(\sigma)$.

In Section~\ref{sec:fairness} we showed that all nodes have equal throughput $\alpha/(1 + \alpha (\beta + 1))$ when using the fair activation rates. When all nodes have equal activation rates, a closed-form expression for the throughput does not seem available. However, we can express the average throughput in terms of the normalization constant $Z_n$.
\begin{proposition}\label{pro:avr_throughput_equal}
Let $\load_i = \sigma$,\ $i = 1,\dots,n$. The average per-node throughput is given by
\begin{equation}\label{eqn:average_throughput}
\avrth_n(\sigma) = \frac{\sigma}{n} \frac{{\rm d} Z_n(\sigma)}{{\rm d} \sigma}\Big/Z_n(\sigma).
\end{equation}
\end{proposition}

\begin{proof}
We have from~\eqref{eqn:throughput_alternative} with $\load_i = \sigma$ that
\begin{equation}
\avrth_n(\sigma) = \frac{\sigma}{n Z_n} \sum_{i = 1}^n Z_{i - \beta -1} Z_{n-i-\beta}.
\end{equation}
We compute, using the definition of $Z_i$ in Theorem~\ref{thm:throughput},
\begin{equation}
\sum_{n = 1}^\infty \Big(\sum_{i = 1}^n Z_{i - \beta -1} Z_{n-i-\beta}\Big)x^n = x\Big(\frac{x^\beta - 1}{x-1} + x^\beta G_Z(x)\Big)^2,
\end{equation}
where $G_Z(x)$ is the generating function of the $Z_i$. Then from~\eqref{eqn:generating_function_Z} and some simplifications, we get
\begin{equation}
\sum_{n = 1}^\infty \big(\sum_{i = 1}^n Z_{i - \beta -1} Z_{n-i-\beta}\big)x^n = \frac{x}{(1 - x - \sigma x^{\beta + 1})^2}.
\end{equation}
On the other hand, we compute from the explicit form~\eqref{eqn:generating_function_Z} of $G_Z(x)$ that
\begin{equation}
\frac{{\rm d}}{{\rm d}\sigma}\big[G_Z(x)\big] = \frac{x}{(1 - x - \sigma x^{\beta + 1})^2},
\end{equation}
and the result follows.
\end{proof}

Using Proposition~\ref{pro:avr_throughput_equal} and the partial fraction expansion~\eqref{dfg2} of $Z_i$, we can express the average per-node throughput in terms of the roots $\lambda_0,\dots\lambda_{\beta}$ of~\eqref{eqn:equation_lambda2}.
\begin{proposition}
Let $\load_i = \sigma$,\ $i = 1,\dots,n$. The average per-node throughput is given by
\begin{equation}\label{eqn:throughput_unfair}
\avrth_n(\sigma) = \frac{\sigma}{n}\frac{P}{Q},
\end{equation}
where
\begin{equation}\label{eqn:throughput_unfair_PQ}
P = \sum_{j = 0}^\beta \frac{\lambda_j^{n+1}}{(\beta + 1) \lambda_j - \beta}\bigg( \frac{n + \beta + 1}{(\beta + 1)\lambda_j - \beta} - \frac{(\beta + 1)\lambda_j}{((\beta + 1) \lambda_j - \beta)^2} \bigg), \quad Q = \sum_{j = 0}^\beta \frac{\lambda_j^{n + \beta + 1}}{(\beta + 1) \lambda_j - \beta}.
\end{equation}
\end{proposition}

\begin{proof}
By~\eqref{dfg} and~\eqref{eqn:c_j} we have
\begin{equation}\label{eqn:proof_throughput_unfair1}
Z_n(\sigma) = \sum_{j = 0}^\beta \frac{\lambda_j^{n + \beta + 1}}{(\beta +1)\lambda_j - \beta},
\end{equation}
where $\lambda_j$ are the $(\beta + 1)$ roots $\lambda$ of~\eqref{eqn:equation_lambda2}. By implicit differentiation of~\eqref{eqn:equation_lambda2} with respect to $\sigma$ we find
\begin{equation}\label{eqn:proof_throughput_unfair3}
\frac{{\rm d} \lambda_j}{{\rm d}\sigma} = \frac{1}{\lambda_j^{\beta - 1}} \frac{1}{(\beta + 1)\lambda_j - \beta}.
\end{equation}
Then from~\eqref{eqn:proof_throughput_unfair1} and~\eqref{eqn:proof_throughput_unfair3} we get
\begin{align}
\nonumber \frac{{\rm d} Z_n(\sigma)}{{\rm d} \sigma} &= \sum_{j = 0}^\beta \bigg( \frac{(n + \beta + 1) \lambda_j^{n + \beta}}{(\beta + 1) \lambda_j - \beta} - \frac{(\beta + 1) \lambda_j^{n + \beta + 1}}{((\beta + 1) \lambda_j - \beta)^2}\bigg) \frac{{\rm d} \lambda_j}{{\rm d} \sigma}\\
&= \sum_{j = 0}^\beta \frac{\lambda_j^{n+1}}{(\beta + 1) \lambda_j - \beta} \bigg( \frac{n + \beta + 1}{(\beta + 1) \lambda_j  - \beta} - \frac{(\beta + 1) \lambda_j}{((\beta + 1) \lambda_j - \beta)^2}\bigg). \label{eqn:proof_throughput_unfair4}
\end{align}
The result follows from substituting~\eqref{eqn:proof_throughput_unfair1} and~\eqref{eqn:proof_throughput_unfair4} into the expression for the average per-node throughput~\eqref{eqn:average_throughput}.
\end{proof}

When the network grows large ($n \rightarrow \infty$) the root of largest modulus, $\lambda_0$, becomes dominant, and~\eqref{eqn:throughput_unfair}-\eqref{eqn:throughput_unfair_PQ} simplifies.
\begin{corollary}\label{col:throughput_unfair_large_n}
Let $\load_i = \sigma$,\ $i = 1,\dots,n$. The limiting average per-node throughput as $n \rightarrow \infty$ is given by
\begin{equation}\label{eqn:throughput_unfair_large_n}
\avrth(\sigma) = \frac{\lambda_0 - 1}{(\beta + 1) \lambda_0 - \beta}.
\end{equation}
\end{corollary}

\begin{proof}
We have, as $n \rightarrow \infty$,
\begin{equation}\label{eqn:throughput_unfair_large_n_2}
P = \frac{\lambda_0^{n+1}}{(\beta + 1)\lambda_0 - \beta} \frac{n}{(\beta + 1) \lambda_0 - \beta} (1 + o(1)), \quad Q = \frac{\lambda_0^{n + \beta + 1}}{(\beta + 1) \lambda_0 - \beta}(1 + o(1)).
\end{equation}
Hence
\begin{equation}
\avrth_n(\sigma) = \frac{\sigma}{n}\frac{P}{Q} = \frac{\sigma \lambda_0^{-\beta}}{(\beta + 1) \lambda_0 - \beta}(1 + o(1)),
\end{equation}
and the result follows as $\sigma \lambda_0^{-\beta} = \lambda_0 - 1$ by~\eqref{eqn:equation_lambda2}.
\end{proof}
The limiting expression~\eqref{eqn:throughput_unfair_large_n} occurs in a variety of contexts in~\cite{PiYe86,Ba2004,ZaMo06,DuDoTh09,JaLeVe09}. When $ \beta \sigma \rightarrow \infty$, the throughput~\eqref{eqn:throughput_unfair_large_n} simplifies even further.
\begin{corollary}\label{col:throughput_unfair_beta_sigma_large}
Let $\load_i = \sigma$,\ $i = 1,\dots,n$ and let $n \rightarrow \infty$. The limiting average throughput as $\beta \sigma \rightarrow \infty$ satisfies
\begin{equation}
\avrth(\sigma) = \frac{1}{\beta + 1}(1 + o(1)).
\end{equation}
\end{corollary}

\begin{proof}
By rewriting~\eqref{eqn:throughput_unfair_large_n} we have
\begin{equation}
\avrth(\sigma) = \frac{1}{\beta + 1} \frac{1}{1 + \frac{1}{(\beta + 1) (\lambda_0 - 1)}}.
\end{equation}
So for $\avrth(\sigma) = \frac{1}{\beta + 1}(1 + o(1))$ to hold, it is necessary and sufficient that $(\beta + 1) (\lambda_0 - 1) \rightarrow \infty$. Recall from~\eqref{eqn:equation_lambda2} that $\lambda_0$ is such that
\begin{equation}
\lambda_0^\beta (\lambda_0 - 1) = \sigma.
\end{equation}
Let $M >0$ be some positive constant, and assume that $\beta \sigma \le M$. Then
\begin{equation}
\beta \lambda_0^\beta (\lambda_0 - 1) = \beta \sigma \le M,
\end{equation}
and so $\beta (\lambda_0 - 1) \le M$. Conversely, assume that $\beta (\lambda_0 - 1) \le K$ for some positive constant $K > 0$. Then
\begin{equation}
\beta \sigma = \beta (\lambda_0 - 1) \lambda_0^\beta  \le \beta (\lambda_0 - 1) {\rm exp}(\beta (\lambda_0 - 1)) \le K {\rm e}^K.
\end{equation}
Hence
\begin{equation}
\beta (\lambda_0 - 1) {\rm ~bounded~} \Leftrightarrow \beta \sigma {\rm ~bounded}.
\end{equation}
It follows that a sufficient condition for $(\beta + 1) (\lambda_0 - 1) \rightarrow \infty$ is that $\beta \sigma \rightarrow \infty$.
\end{proof}
Corollary~\ref{col:throughput_unfair_beta_sigma_large} implies for $\beta$ fixed and $\sigma \rightarrow \infty$ that $\avrth(\sigma) \rightarrow \frac{1}{\beta + 1}$. So for $n \rightarrow \infty$, both the equal and fair activation rates can achieve the maximum throughput by letting $\sigma\rightarrow \infty$ and $\alpha \rightarrow \infty$, respectively. This can be explained by the observation that for both sets of activation rates, the system spends almost all the time in maximal independent sets of active nodes, thus maximizing spatial reuse.

Next, fix $\sigma >0$ and search for $\alpha = \alpha_n(\sigma)$ such that
$$
\avrth_n(\sigma)  =\frac{\alpha}{1 + \alpha(\beta+1)},
$$
so the network-average throughput is identical for the fair rates and equal rates. For $\alpha(\sigma) := \lim_{n \rightarrow \infty} \alpha_n(\sigma)$ we can make this comparison explicit.

\begin{proposition}
We have
\begin{equation}\label{eqn:alpha(sigma)}
\alpha(\sigma) = \lambda_0 - 1.
\end{equation}
\end{proposition}
\begin{proof}
This follows at once by equating~\eqref{eqn:throughput_fair} and~\eqref{eqn:throughput_unfair_large_n} and solving for $\alpha$.
\end{proof}

\begin{figure}[h]
 \begin{center}
    \includegraphics[width = 0.6\linewidth]{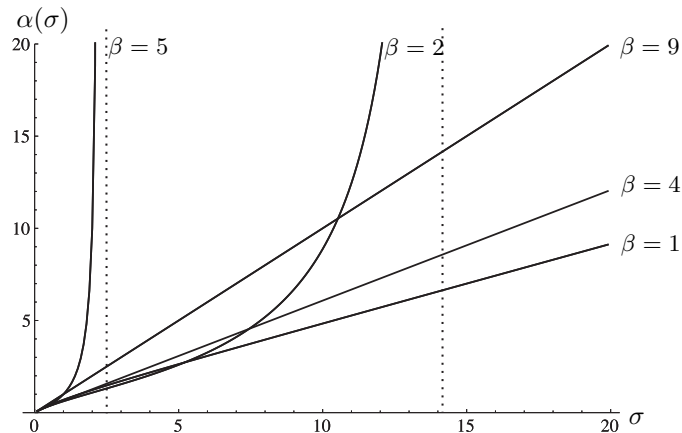}
 \end{center}
 \caption{$\alpha_n(\sigma)$ plotted against $\sigma$ for various values of $\beta$ and $n = 10$}
 \label{fig:aggregate_throughput_equal}
\end{figure}

It is intuitively clear that imposing fairness may compromise the throughput. From~\eqref{eqn:throughput_fair} it is seen that the fair per-node throughputs are bounded above by $\frac{1}{\beta + 1}$, and that this upper bound can be approached by letting $\alpha \rightarrow \infty$. Corollary~\ref{col:throughput_unfair_large_n} shows that, as $n \rightarrow \infty$, the average throughputs in the fair case and unfair case are equal when $\alpha$ is taken to be $\lambda_0 - 1$. The maximum activation rate in this limiting case equals
\begin{equation}
\fsolv_{\rm max} = \alpha(1 + \alpha)^\beta = \load_i, \quad \beta + 1\le i \le n - \beta
\end{equation}
as is seen from~\eqref{eqn:fair_load}. This maximum grows like $\sigma$ since $1 + \alpha = \lambda_0 = \sigma^{1/(\beta + 1)}(1 + o(1))$,\ $\sigma \rightarrow \infty$, according to~\eqref{eqn:series_expansion_large}. Hence, as $n \rightarrow \infty$, the fair case achieves the same average throughput with activation rates that are of smaller or of the same size as as in the unfair case. The situation is considerably more complicated when we keep $n$ bounded. Then it may well occur that $\avrth_n(\sigma)$ exceeds $\frac{1}{\beta + 1}$, which is the upper bound for the throughput achievable by the fair scheme. In Figure~\ref{fig:aggregate_throughput_equal} we have plotted curves $\alpha = \alpha_n(\sigma)$, for the case that $n = 10$ and various values of $\beta$ in the range $0 \le \sigma \le 20$. The occurrence of the asymptotes for the curves with $\beta = 2,5$ shows that $\avrth_n(\sigma) > \frac{1}{\beta + 1}$ when $\sigma$ is to the right of the asymptote. Remarkably, in the cases that $\beta = 1,4,9$, no such asymptotes exist. This can be explained by the observation that a system operating with equal rates $\sigma \rightarrow \infty$ has at all times $\lceil \frac{n}{\beta + 1}\rceil$ nodes activated, cf.~\cite{DeBoVeHi08}. When $\frac{n}{\beta + 1}$ is integer, the resulting average throughput does not exceed the upper bound $\frac{1}{\beta + 1}$ for the throughput in the fair case and so no asymptote occurs.

\section{Non-saturated networks}\label{sec:non-saturated}

The saturated model described in Section~\ref{sec:model} assumes all nodes to have an infinite supply of packets pending for transmission. We now relax this assumption by introducing queueing dynamics. Packets arrive at node~1 according to a Poisson process with rate~$\ar$, and are routed along nodes $2,\dots,n$. When node~$n$ finishes a transmission, the corresponding packet exits the system. All nodes have infinite buffer capacity. These dynamics describe a non-saturated network, where packets are forwarded through the network, using nodes $2,\dots,n$ as relays. As nodes are no longer saturated, they may not always have packets available for transmission. When the backoff timer of a node expires, and the node is not blocked, it can activate if its buffer contains at least one packet. Otherwise, the node will remain inactive and restart its backoff timer.

To describe the state of the non-saturated network we have to take into account the queue length at each node, as well as node activity. The stochastic process that describes both the activity of nodes and their queue lengths is once again a Markov process, but does not have the appealing product-form limiting distribution encountered in the saturated network. The added complexity of the model is reflected by the lack of available analytical results. When node~1 is saturated (i.e. $\ar \rightarrow \infty$) and all nodes have equal activation rate $\sigma \rightarrow \infty$ the system simplifies, and in some cases can be analyzed. For $n = 3$, $\beta = 1$ it can be shown (cf.~\cite{DeBoVeHi08}) that both nodes~1 and~2 are unstable, so that the number of packets in their buffers grows without bounds. Using this,~\cite{DeBoVeHi08} shows the throughput of node~3 to be equal to $3/10$. Shneer and Van de Ven~\cite{ShVe09} demonstrate that for general $\beta$ and $n = 2 \beta + 1$, nodes $1,\dots, \beta + 1$ are unstable, and the throughput of node~$n$ is bounded as $1/(2 \beta + 2) \le \theta_n \le 1/(2 \beta + 1)$.

The throughput of node~$n$ is of special interest, as it represents the end-to-end throughput of the network, that is, the intensity at which packets leave the network. When $\theta_n = \ar$, all packets that enter the network eventually leave. If $\theta_n < \ar$, packets arrive into the system at a higher rate than the network can process, and packets will accumulate at certain bottleneck nodes. The two results for the special case in~\cite{DeBoVeHi08} and~\cite{ShVe09} paint a particularly bleak picture as, despite the high activation rate, the end-to-end throughput is well below the theoretical upper bound $1/(\beta +1)$ that can be attained through centralized schedulers. Instead, we investigate the performance of a non-saturated network that uses the fair rates~\eqref{eqn:fair_load}.

\begin{figure}[h]
 \begin{center}
    \includegraphics[width = 0.5\linewidth]{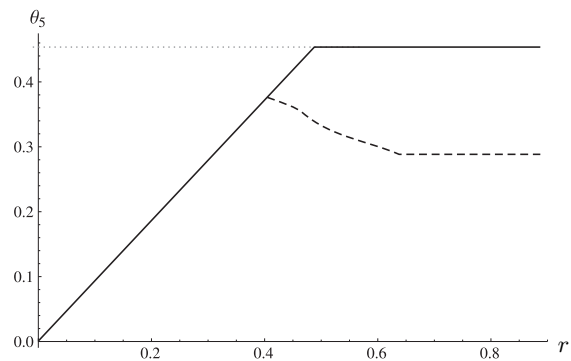}
 \end{center}
 \caption{The end-to-end throughput of a network with equal activation rates (dashed) and fair activation rates (solid), plotted against the arrival rate at node~1}
 \label{fig:multihop_end-to-end_vs_input}
\end{figure}

Figure~\ref{fig:multihop_end-to-end_vs_input} shows simulation results for the end-to-end throughput in a network using the fair rates, with $n = 5$ and $\beta = 1$, plotted against the arrival rate~$\ar$. The dashed line corresponds to the network where all nodes have equal activation rate $\sigma  = 6$, and the solid line shows the throughput of a network with fair activation rates~\eqref{eqn:fair_load} and $\alpha = 11.68$. The values of $\sigma$ and $\alpha$ are chosen such that the average per-node throughput is equal for both activation schemes, see Section~\ref{sec:throughput}. The network with equal activation rates performs poorly. When the arrival rate grows beyond a certain threshold, node~2 saturates and the throughput drops. The network with fair rates, on the other hand, can sustain higher arrival rates and experiences no throughput drop when in overload. In fact, the end-to-end throughput approaches the per-node throughput in the corresponding saturated network (indicated by the dotted horizontal line). So the network is stable whenever $\ar < \alpha/(1 + \alpha(\beta + 1))$.

\begin{figure}[h!]
 \begin{center}
 \subfigure[Equal activation rates]{\label{fig:activitychart_equal} \includegraphics[width= 0.45\linewidth]{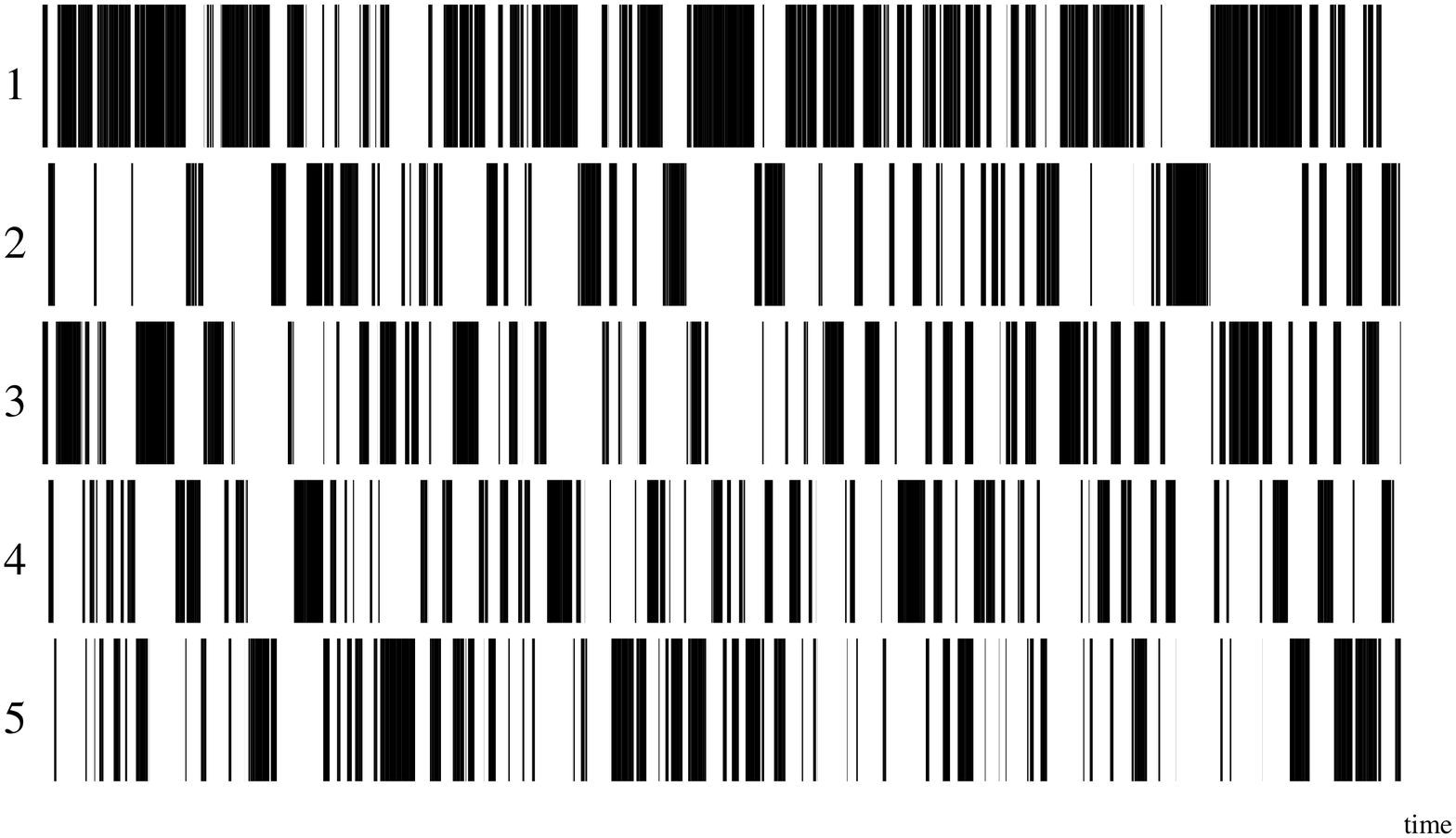}} \hspace{0.5cm}
 \subfigure[Fair activation rates]{\label{fig:activitychart_fair} \includegraphics[width= 0.45\linewidth]{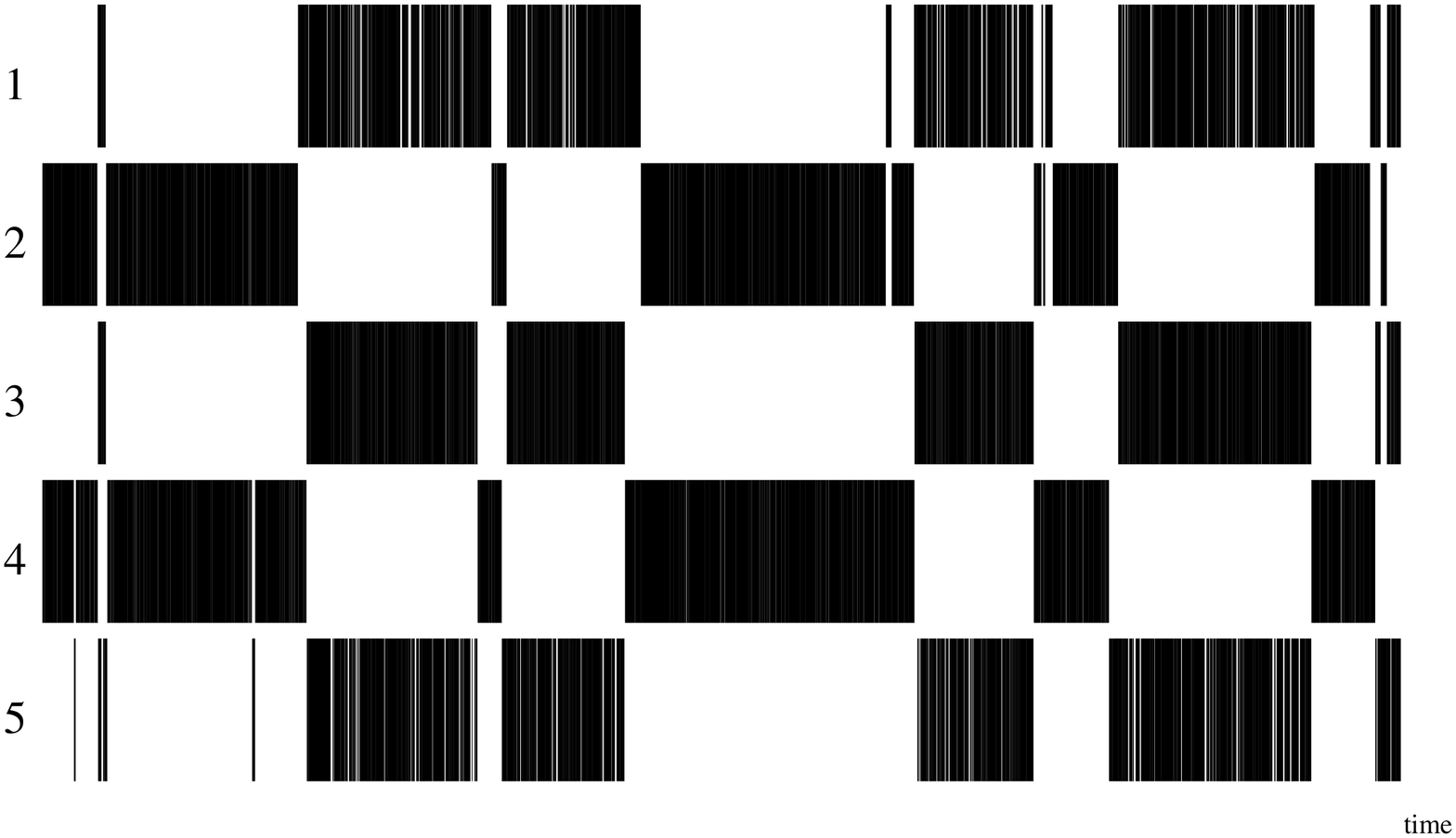}}
 \end{center}
  \caption{The activity charts for both equal activation rates and fair activation rates for $n = 5$, $\beta = 1$ and $\ar = 0.47$}
 \label{fig:activity_chart}
\end{figure}

We see from Figure~\ref{fig:multihop_end-to-end_vs_input} that as $\ar$ increases, the end-to-end throughput eventually reaches the per-node throughput in the saturated network for $\alpha = 11.68$. Sustaining this high end-to-end throughput requires an efficient activation scheme, in which maximal independent sets of active nodes are persistant for a long time, yet all nodes get equal access. Figure~\ref{fig:activity_chart} shows a typical sample of the activity pattern, for both the equal rates and the fair rates. The arrival rate equals $\ar$ equals 0.47, so the network with fair rates is stable. The barcode plots indicate for each point in time whether nodes are active (black) or inactive (white). We see from Figure~\ref{fig:activitychart_equal} that for equal activation rates, the pattern is very irregular. Node~1 is more active than the other nodes, indicating that the network is in fact unstable, and node~2 is the bottleneck node. The fair activation rates result in a more regular pattern, and maintain certain maximal independent sets of active nodes for a long period of time. Figure~\ref{fig:activitychart_fair} shows that the maximal independent sets $\{1,3,5\}$ and $\{2,4\}$ are persistent. So instead of sending a single packet before relinquishing access to the medium, nodes send packets in large batches.

\begin{figure}[h!]
 \begin{center}
 \subfigure[Node~1]{\label{fig:queue_node_1_largescale} \includegraphics[width= 0.3\linewidth]{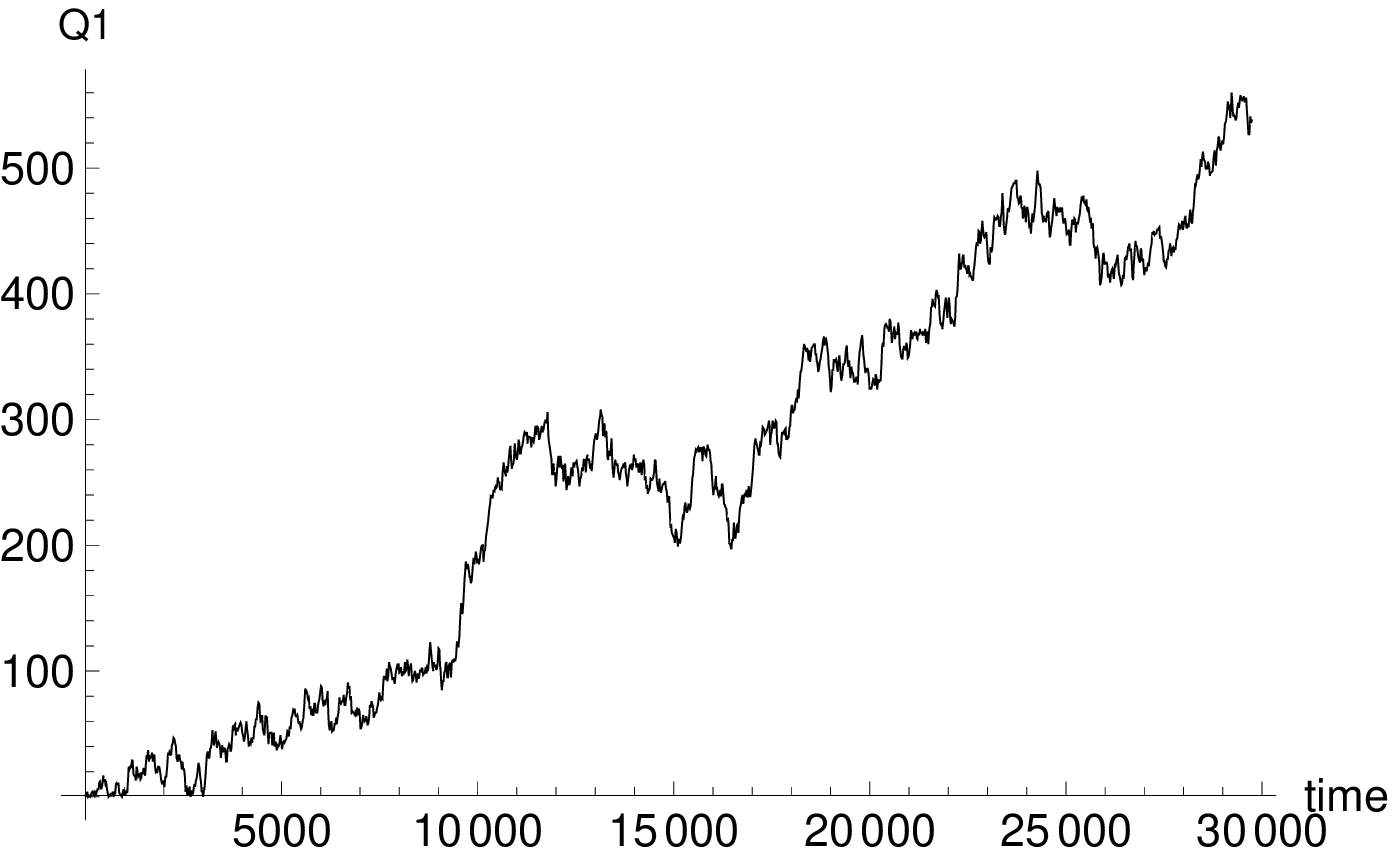}}
 \subfigure[Node~2]{\label{fig:queue_node_2_largescale} \includegraphics[width= 0.3\linewidth]{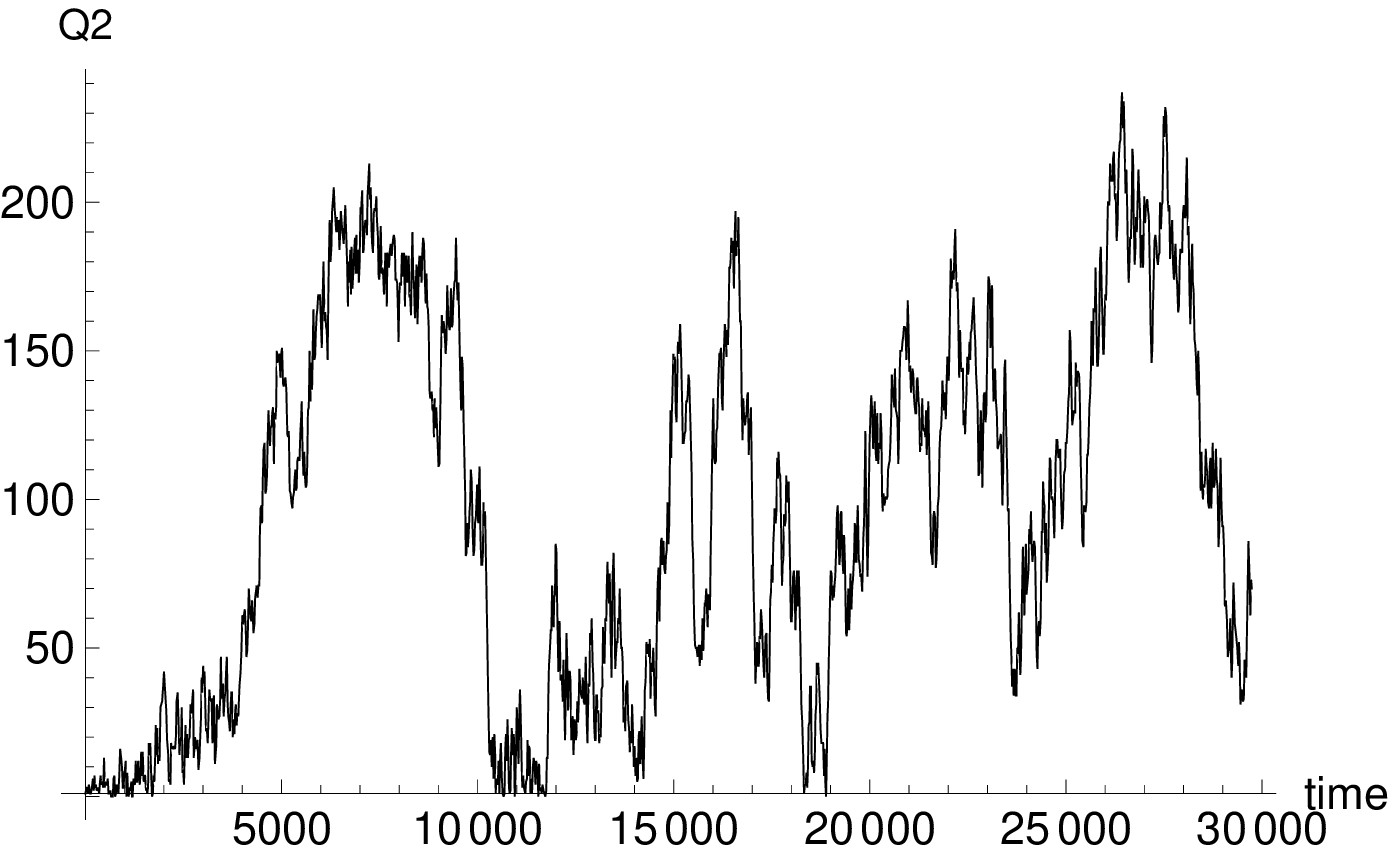}}
 \subfigure[Node~3]{\label{fig:queue_node_3_largescale} \includegraphics[width= 0.3\linewidth]{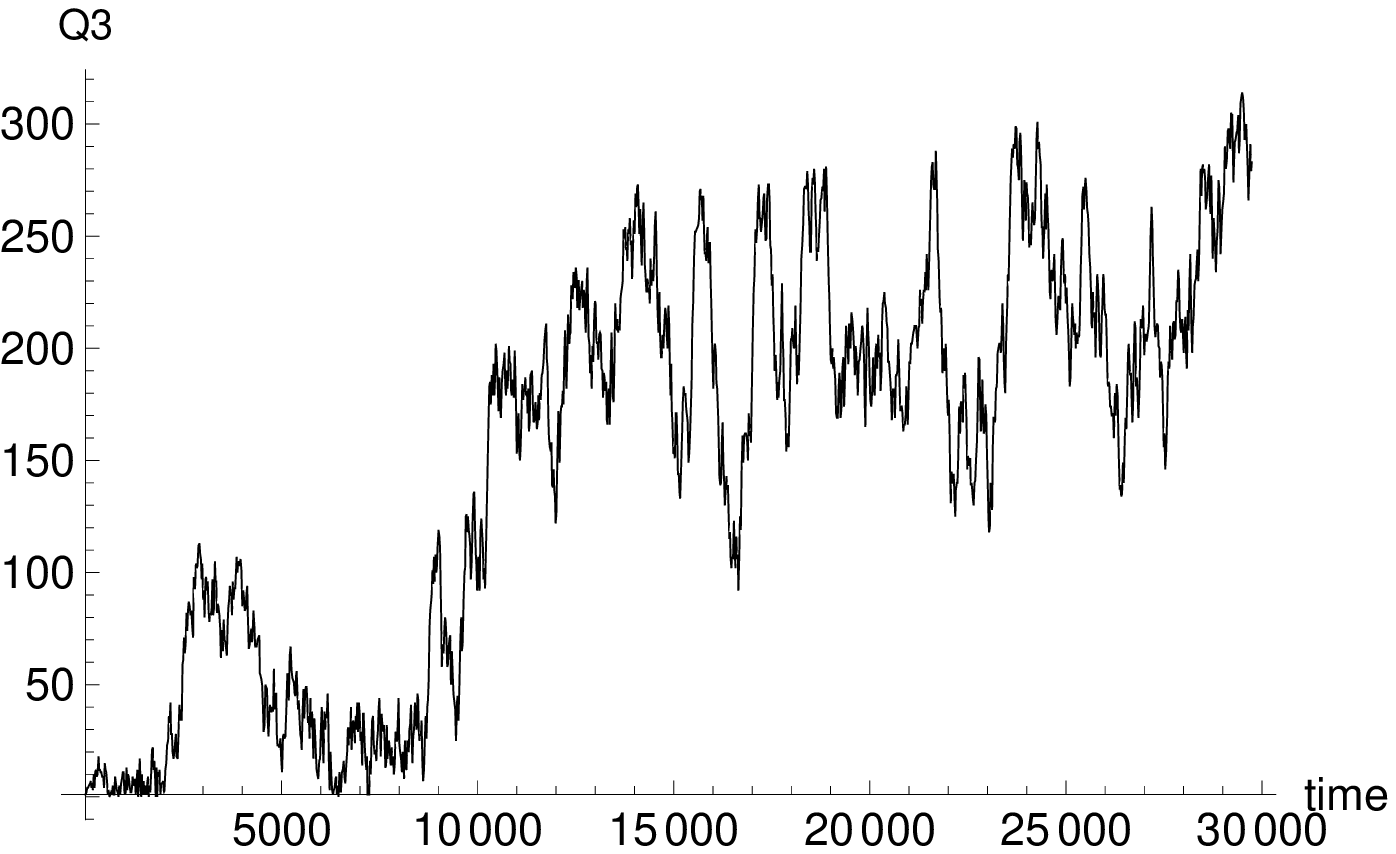}}
 \subfigure[Node~4]{\label{fig:queue_node_4_largescale} \includegraphics[width= 0.3\linewidth]{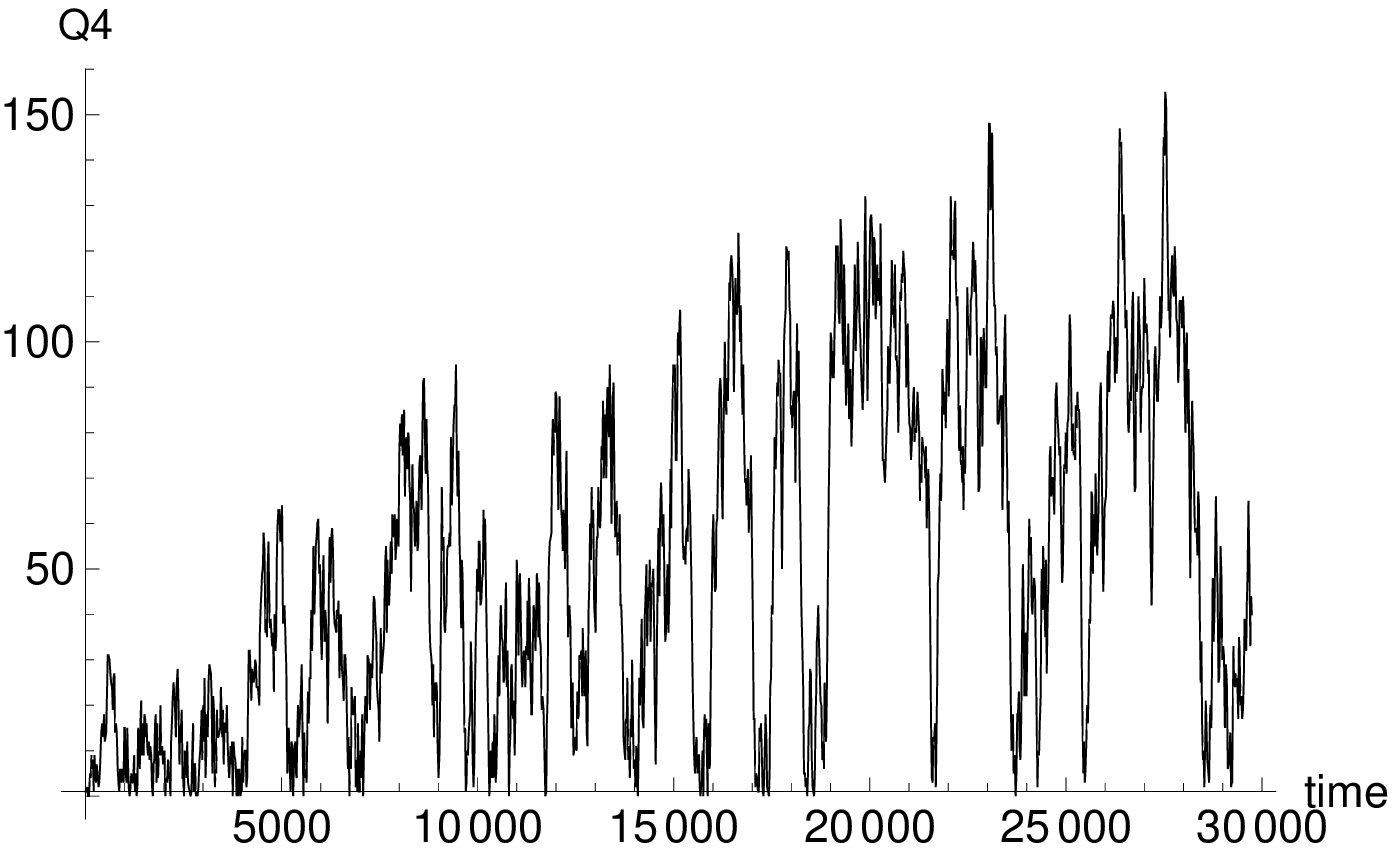}}
 \subfigure[Node~5]{\label{fig:queue_node_5_largescale} \includegraphics[width= 0.3\linewidth]{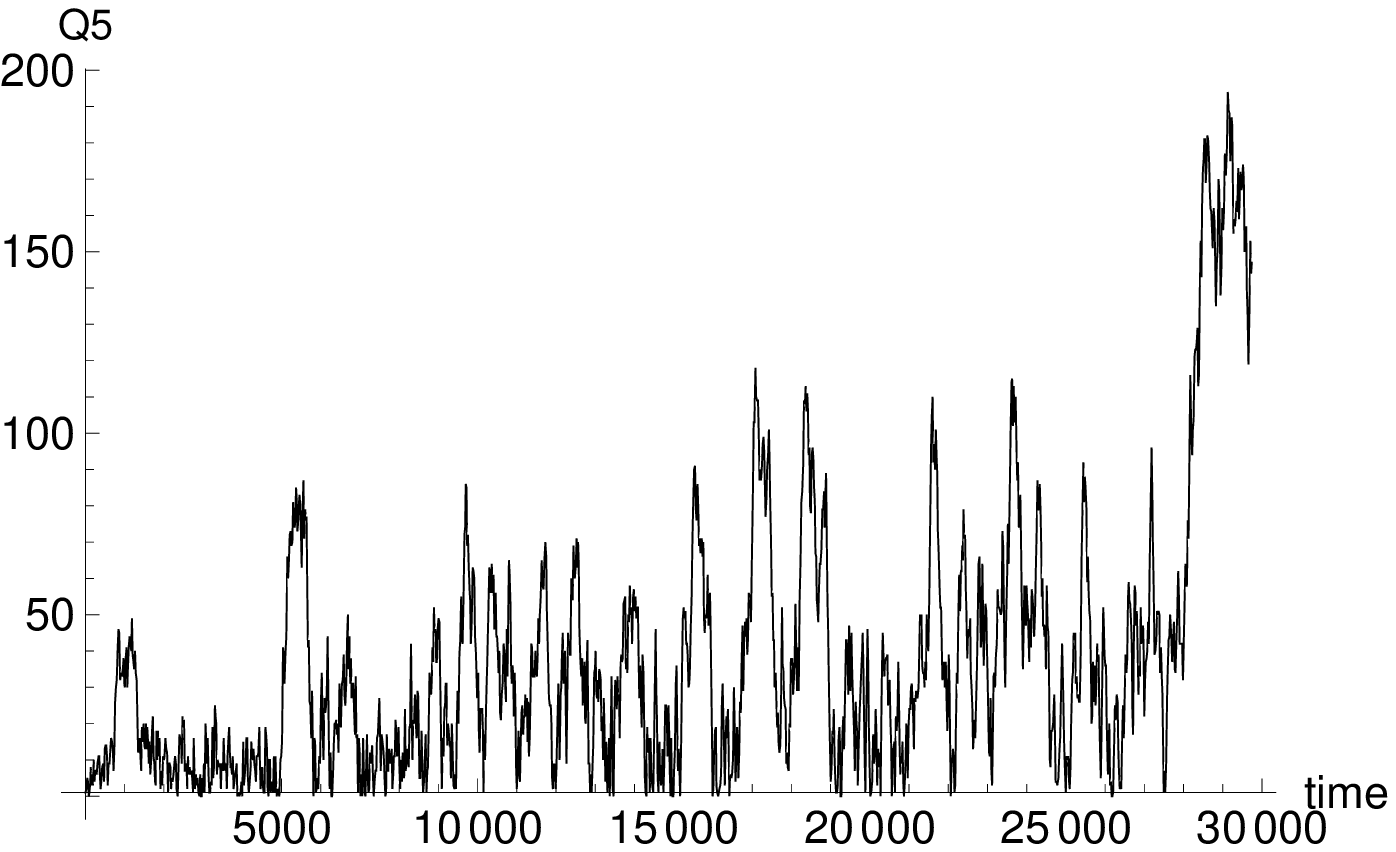}}
 \end{center}
  \caption{The queue lengths for the various nodes with fair activation rates, $n = 5$, $\beta = 1$ and $\ar = 0.47$, on a large time scale}
 \label{fig:queue_lengths_largescale}
\end{figure}

The length of the dominant periods of a maximal independent set of active nodes, and therefore the size of the batches, is related to the queue lengths of the nodes in this set. Typically, when one of the queues of the dominant maximal independent set empties, dominance reverts to the other maximal independent set. Figure~\ref{fig:queue_lengths_largescale} shows the queue lengths of all nodes over time. It shows large fluctuations in queue length, which are directly related to the large batch sizes. Nodes~4 and~5 in particular empty relatively often, suggesting that these are in fact the nodes that typically break the dominance of their own maximal independent set.


\begin{figure}[h!]
 \begin{center}
 \subfigure[Node~1]{\label{fig:queue_node_1_smallscale} \includegraphics[width= 0.3\linewidth]{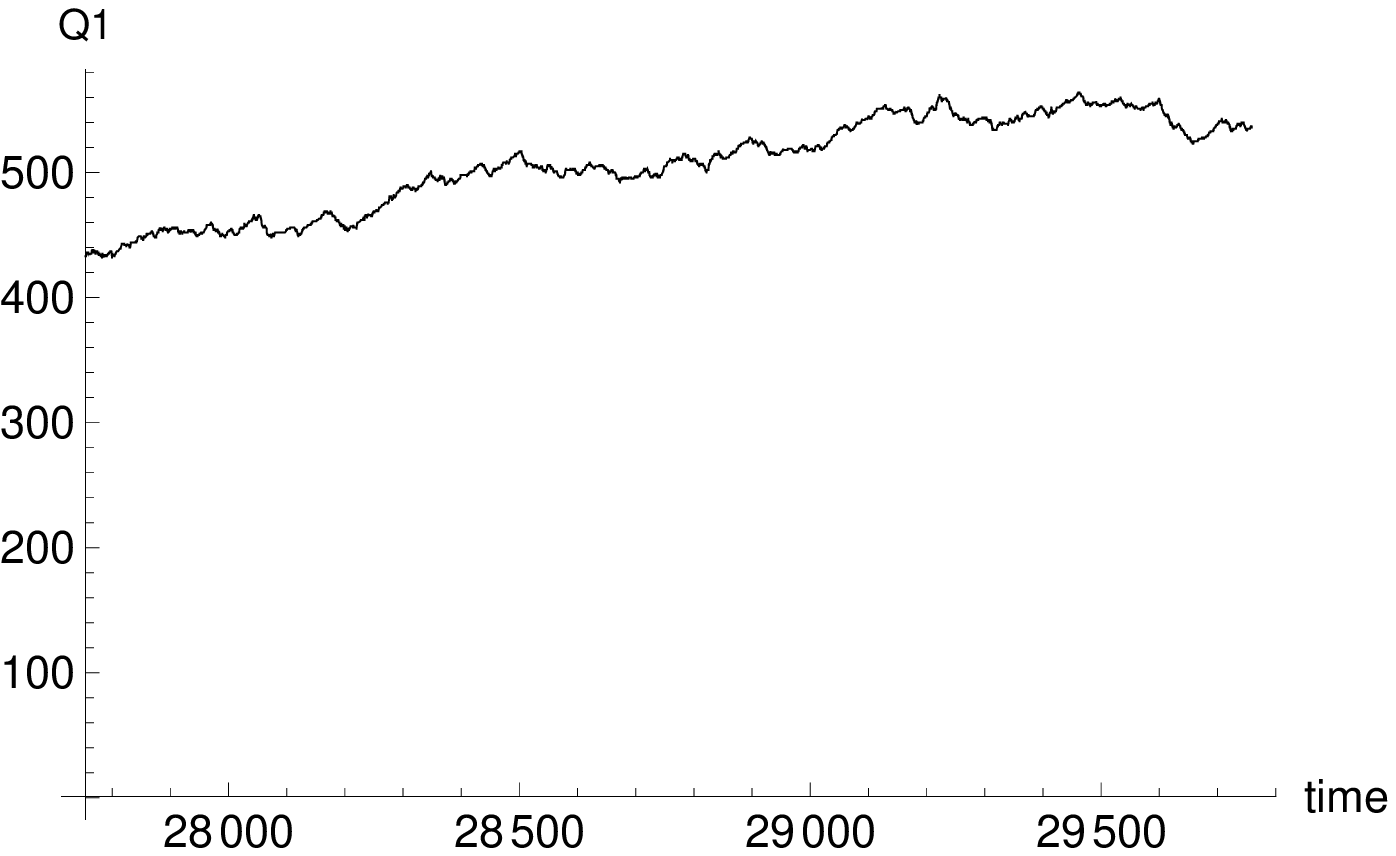}}
 \subfigure[Node~2]{\label{fig:queue_node_2_smallscale} \includegraphics[width= 0.3\linewidth]{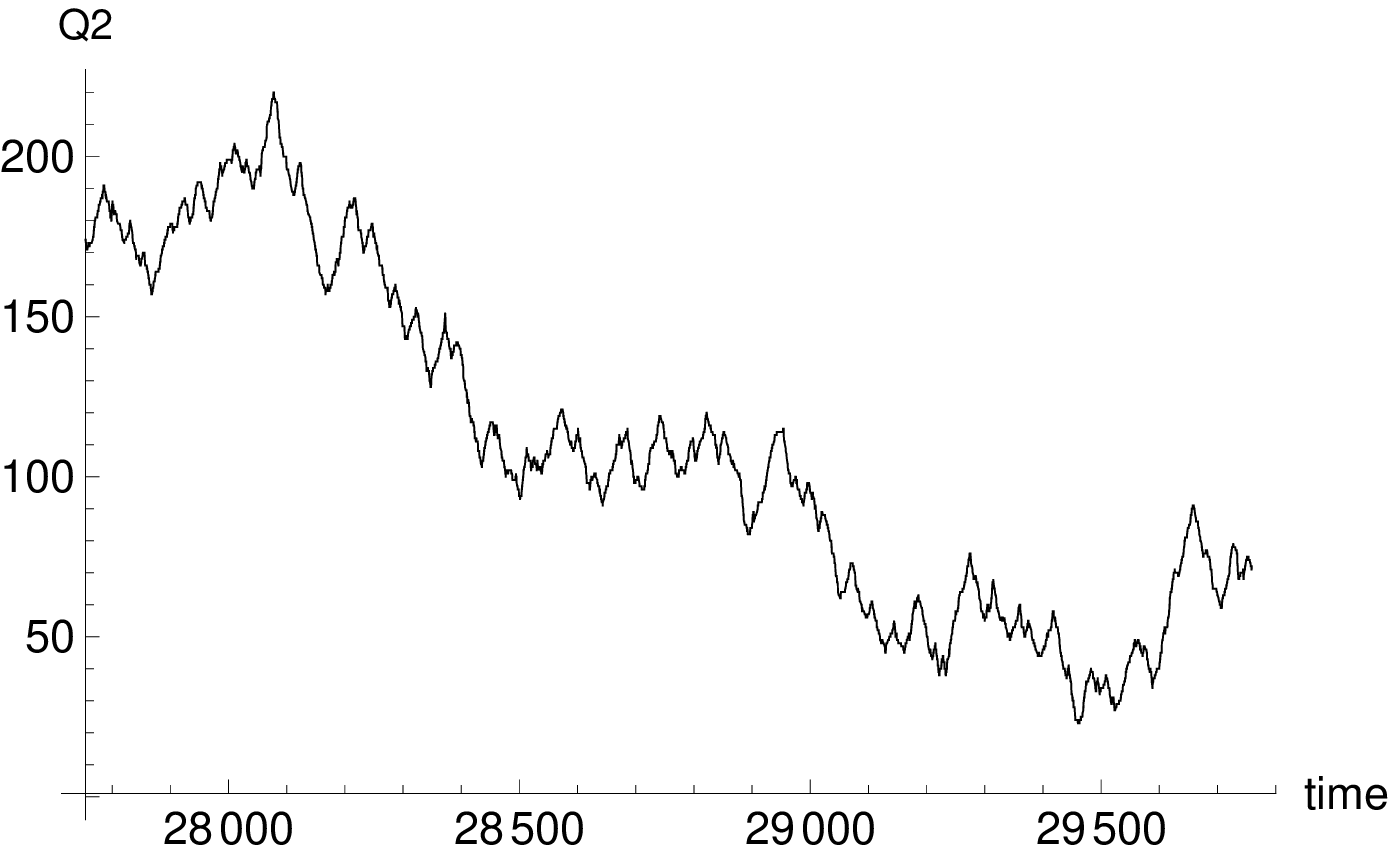}}
 \subfigure[Node~3]{\label{fig:queue_node_3_smallscale} \includegraphics[width= 0.3\linewidth]{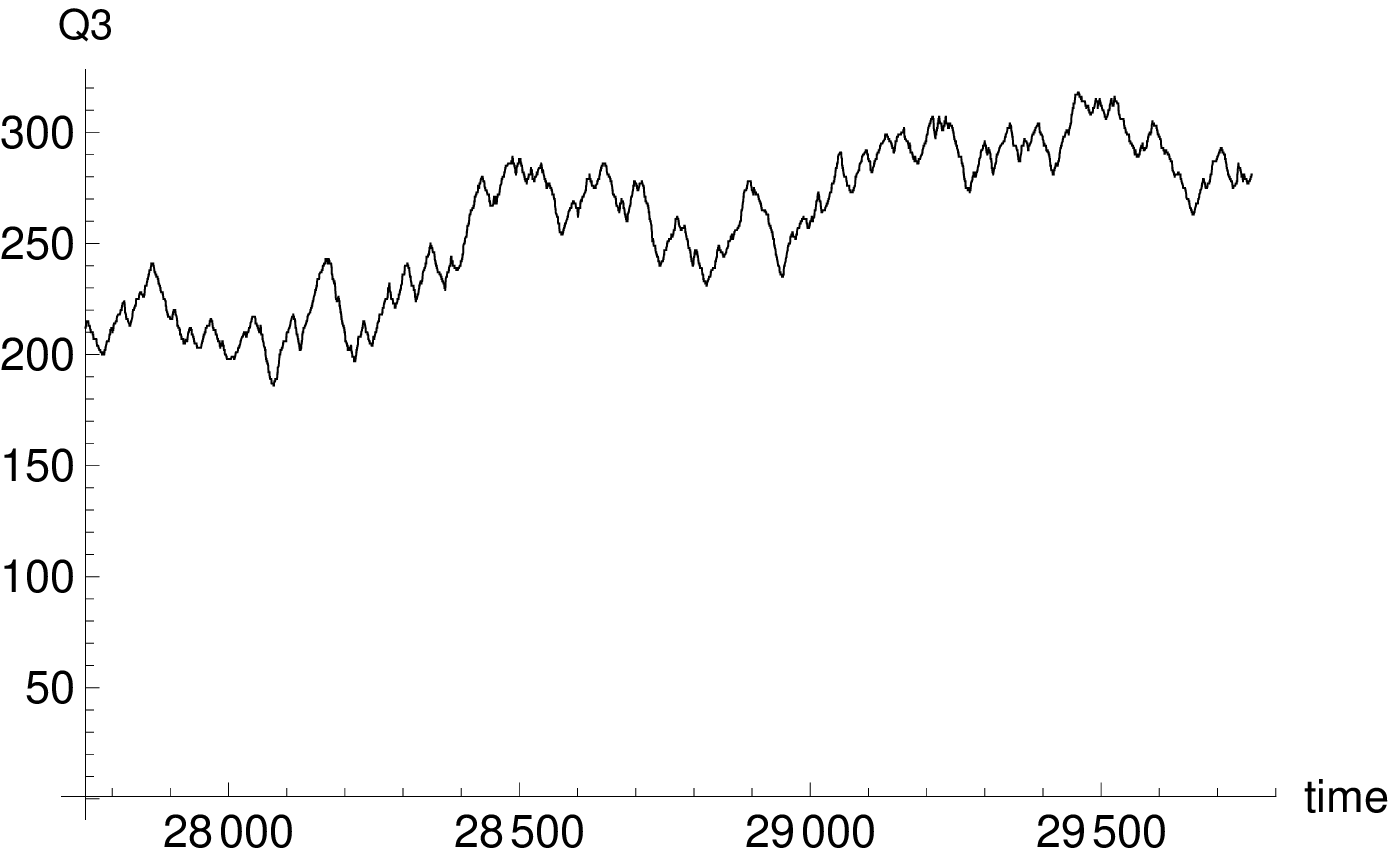}}
 \subfigure[Node~4]{\label{fig:queue_node_4_smallscale} \includegraphics[width= 0.3\linewidth]{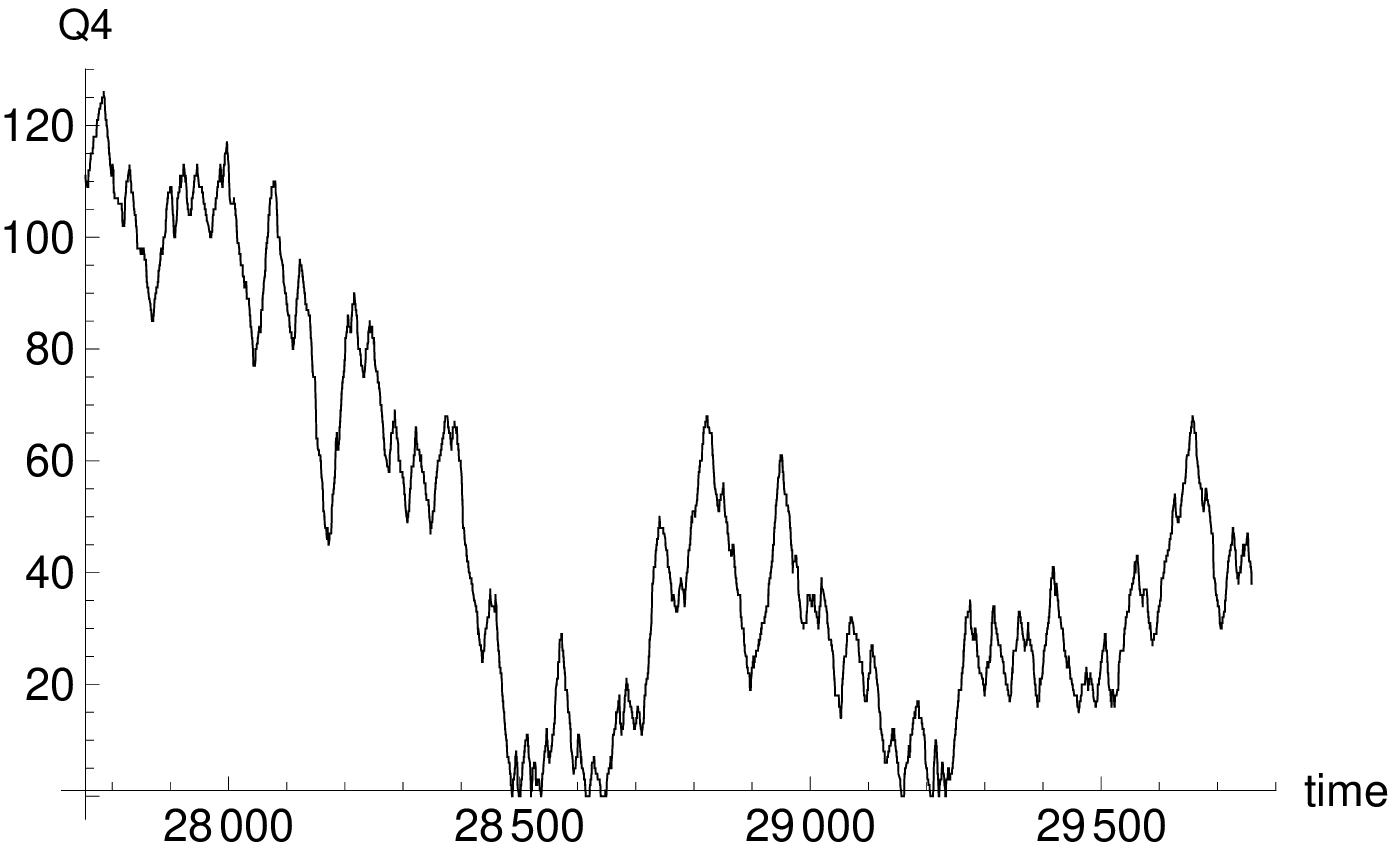}}
 \subfigure[Node~5]{\label{fig:queue_node_5_smallscale} \includegraphics[width= 0.3\linewidth]{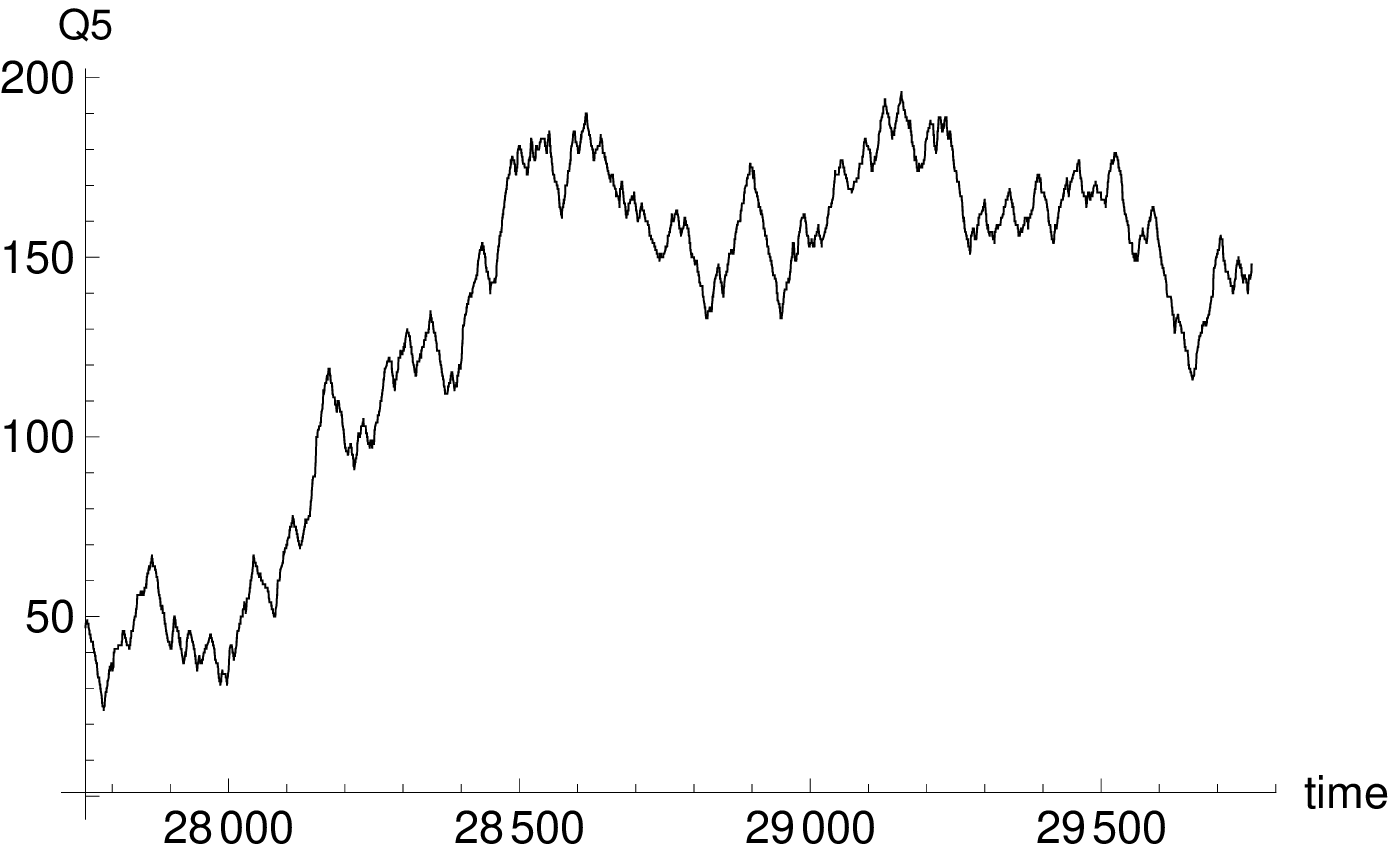}}
 \end{center}
  \caption{The queue lengths for the various nodes with fair activation rates, $n = 5$, $\beta = 1$ and $\ar = 0.47$, on a small time scale}
 \label{fig:queue_lengths_smallscale}
\end{figure}

In contrast to the saturated network, the non-saturated network may waste capacity when unblocked nodes cannot activate because of empty buffers. Figure~\ref{fig:queue_lengths_smallscale} is similar to Figure~\ref{fig:queue_lengths_largescale}, except for the smaller time scale. It suggests that queues rarely empty, and moreover do not stay empty for a long time, so the wasted capacity is very small, if not negligible.

The results presented in this section carry over to general $n$ and $\beta$. The system with fair activation rates can sustain arrival rates up to the per-node throughput of the corresponding saturated network, and maintains its throughput when $r$ grows even larger. The network with equal activation rates with the same (saturated) average per-node throughput fails to attain this end-to-end throughput. Whenever this system is overloaded, its throughput drops. Recall from~\eqref{eqn:throughput_fair} that for $\alpha \rightarrow \infty$, the saturated network approaches a per-node throughput of $1/(\beta + 1)$. So the corresponding non-saturated network even approaches the theoretic upper bound on the end-to-end throughput.

The high end-to-end throughput of a non-saturated network using the fair rates, and the poor performance of such a network where all nodes have equal activation rates, can be explained by the result derived for saturated networks in Sections~\ref{sec:unfairness} and~\ref{sec:fairness}. Indeed, although non-saturated networks are more complex than their saturated counterparts, the basic premise that nodes in the middle of the network have an unfavorable position compared with nodes on the network boundary remains.

\section{Conclusions and outlook}\label{sec:conclusions}

In this paper we studied the unfairness in linear multi-access networks. We proposed node-specific activation rates in~\eqref{eqn:fair_load} as a function of the number of neighbors, and showed that these rates provide equal throughput for all nodes. The rates increase with the number of neighbors. Intuitively, this structure can be explained by the observation that, as the number of neighbors increases, a node needs a higher activation rate to retain its throughput. Consequently, the rule in~\eqref{eqn:fair_load}, which is exact in linear networks, might serve as a heuristic in more complex networks. 

Finding exact expressions for the activation rates that provide strict fairness for networks beyond the linear network is challenging. Kelly~\cite{Kelly85} obtained results for trees with nearest-neighbor blocking, and finds that rates such as in~\eqref{eqn:fair_load}, where nodes on the leafs of the tree have lower rates than those in the stem of the tree, provide strict fairness. For such trees, it seems possible to extend Kelly's result to the $\beta$-hop blocking situation. Also other regular topologies such as certain grids appear to admit similar analysis.

Finally, let us mention that regular networks like grids or trees may not always be a good representation of topologies encountered in practice, which in general are less structured. The results obtained in this paper, however, rely heavily on the diagonal structure of the capacity matrix $A$ in~\eqref{eqn:Matrix_A}, which only exists for certain well-structured networks. For more general networks, and hence more general matrices $A$, the objective of equal througputs boils down to solving the system of nonlinear equations that follows from \eqref{eqn:throughput}. In fact, \eqref{eqn:throughput} can be described in terms of the
the mapping (with $\load = (\load_1,\dots,\load_n) \in (0,\infty)^n$)
\begin{equation*}
\load  \mapsto \theta(\load) = Z_n \cdot X \cdot \Pi = \Bigg( \sum_{\substack{\omega \in \Omega\\ \omega_k \neq 0}} \prod_{i = 1}^n \load_i ^{\omega_i }\Bigg)_{k = 1,\dots,n}
\end{equation*}
with $\theta(\load) \in (0,\infty)^n$.
It can be shown that the mapping $\theta$ is globally invertible on $(0,\infty)^n$. Thus, given a vector $c \in (0,\infty)^n$, there is a unique $\load = \load(c) \in (0,\infty)^n$ such that $\theta(\load) = c$. When $c$ has identical entries, this corresponds to all nodes having equal throughput. The full analysis of this fixed-point equation is rather involved and will appear elsewhere.

\appendix

\section{Normalization constant}\label{sec:partition_function}

We take a closer look at the normalization constant $Z_i$ defined in~\eqref{eqn:def_Z_small}-\eqref{eqn:def_Z_large}, in the regime where all nodes have equal activation rates $\load_i = \sigma$. This section closely follows Pinksy and Yemini~\cite{PiYe86} and Janssen {\em et al.}~\cite{JaLeVe09}. All proofs can be found in~\cite{JaLeVe09}.

The generating function $G_Z(x)$ of the $Z_i$ is given by
\begin{equation}\label{eqn:generating_function_Z}
G_Z(x) = \sum_{i=0}^\infty Z_i x^i = \frac{x-1 + \sigma x^{\beta+1} - \sigma x}{(x-1) (1 - x - \sigma x^{\beta+1})}.
\end{equation}
Let $\lambda_0,\dots,\lambda_{\beta}$ denote the $\beta+1$ roots of
\begin{equation}\label{eqn:equation_lambda}
\lambda^{\beta+1} - \lambda^\beta - \sigma = 0.
\end{equation}
These roots can be shown to be distinct, and from Rouch\'e's theorem it follows that there exists a unique positive real root $\lambda_0$ such that $\lambda_0 >|\lambda_j|,\ j = 1,\dots,\beta$. We can obtain the $Z_i$ from the generating function by applying partial fraction expansion, which gives (\cite{JaLeVe09}, Proposition 1)
\begin{equation}\label{dfg}
Z_i = \sum_{j = 0}^\beta c_j \lambda_j^i \quad , i =0,1,\dots,
\end{equation}
where $\lambda_j$ are the roots of \eqref{eqn:equation_lambda}, and
\begin{equation}\label{eqn:c_j}
c_j = \frac{\lambda_j^{\beta + 1}}{(\beta+1)\lambda_j - \beta}.
\end{equation}
When the index grows large ($i \rightarrow \infty$), the normalization constant~\eqref{dfg} is dominated by the largest root $\lambda_0$, thus simplifying considerably, i.e.,
\begin{equation}\label{eqn:partition_function_large}
Z_i = c_0 \lambda_0^i\left(1 + o(1)\right), \quad i \rightarrow \infty.
\end{equation}

The roots $\lambda_0,\dots,\lambda_\beta$ have representations in terms of infinite series expressions. Let $(x)_n = \Gamma(x+n)/\Gamma(x)$ denote the Pochhammer symbol. Depending on the value of $\sigma$, either of the following two expansions applies (See \cite{JaLeVe09}, Proposition 7 and 8). For small $\sigma > 0$, we have
\begin{align}
\lambda_0(\sigma) &= 1 + \sum_{l = 1}^\infty \frac{(-1)^{l-1} (\beta l)_{l-1}}{l!} \sigma^l, \label{eqn:series_expansion_small_0}\\
\lambda_j(\sigma) &= \sum_{l = 1}^\infty \frac{(l/\beta)_{l-1}}{l!} w_j^l, \quad j = 1,2,\dots,\beta, \label{eqn:series_expansion_small_j}
\end{align}
where $w_j = \sigma^{1/\beta} {\rm e}^{2 \pi \imath (j - 1/2)/\beta}$. The series expansions in~\eqref{eqn:series_expansion_small_0} and~\eqref{eqn:series_expansion_small_j} converge for
\begin{equation*}
0 \le \sigma \le \frac{\beta^\beta}{(\beta+1)^{\beta+1}}=:\xi(\beta),
\end{equation*}
and diverge otherwise. For large $\sigma > 0$, we have
\begin{equation}
\lambda_j(\sigma)^{-1} = \sum_{l = 1}^\infty \frac{\left(\frac{-l}{\beta+1}\right)_{l-1}}{l!} v_j^{-l}, \label{eqn:series_expansion_large}
\end{equation}
where $v_j = \sigma^{1/(\beta+1)} {\rm e}^{2 \pi \imath j/(\beta+1)}$. The series expansion in~\eqref{eqn:series_expansion_large} converges for $\sigma \ge \xi(\beta)$, and diverges otherwise.

%

\section{Proof of Proposition 1}\label{app1}

We first establish an auxiliary result. Define $\abc(i,l,n)$ as the number of states in which exactly $l$ nodes are active, including node $i$. For successive nodes, the following relation holds.

\begin{lemma}\label{lem:single_basic}
Let $n \in \mathds{N}, i \leq \left\lceil\frac{n}{2}\right\rceil - 1$. Then:
\begin{align}
\abc(i,l,n) &= \abc(i+1,l,n),~~~l \leq i,\label{eqn:single_basic_2}\\
\abc(i,l,n) &> \abc(i+1,l,n),~~~i~{\rm odd},~i < l \leq \left\lceil n/2\right\rceil, \label{eqn:single_basic_1}\\
\abc(i,l,n) &< \abc(i+1,l,n),~~~i~{\rm even},~i < l \leq \left\lceil n/2\right\rceil. \label{eqn:single_basic_3}
\end{align}
\end{lemma}

\begin{proof}
The proof is by induction on $i$. Conditioning on activity of node~1 and node~$n$ yields the relations
\begin{align}
\abc(i,l,n) &= \abc(i-2,l-1,n-2) + \abc(i-1,l,n-1), \label{eqn:recursion_forw}\\
\abc(i,l,n) &= \abc(i,l-1,n-2) + \abc(i,l,n-1), \label{eqn:recursion_backw}
\end{align}
with boundary conditions
\begin{align*}
\abc(0,l,n) &= 0 {\rm ~for~all~} n {\rm ~and~} l;\\
\abc(1,l,n) &= 1 {\rm~for~} l > 0 {\rm ~and~all~} n;\\
\abc(1,l,n) &= 0 {\rm~for~} l \leq 0 {\rm ~and~all~} n.
\end{align*}
Hence, the initialization step of the induction is
\begin{align*}
\abc(0,l,n) &< \abc(1,l,n),~~~0 < l < \left\lceil n/2 \right\rceil,\\
\abc(0,l,n) &= \abc(1,l,n),~~~l \leq 0.
\end{align*}
Consider odd $i \leq \left\lceil n/2 \right \rceil - 2$, let $i+1 < l < \left\lceil n/2\right\rceil$, and assume $\abc(i,l,n) > \abc(i+1,l,n)$. Using \eqref{eqn:recursion_forw} and \eqref{eqn:recursion_backw} we get
\begin{align*}
\abc(i+1,l,n) &= \abc(i+1,l-1,n-2) + \abc(i+1,l,n-1)\\
&< \abc(i, l-1,n-2) + \abc(i+1,l,n-1) = \abc(i+2,l,n).
\end{align*}
This proves assertion~\eqref{eqn:single_basic_1}. Assertions~\eqref{eqn:single_basic_2} and~\eqref{eqn:single_basic_3} can be proved in a similar manner.
\end{proof}

We now use Lemma~\ref{lem:single_basic} to prove Proposition~\ref{pro:single_k=1}.

\begin{proof}{\rm (Proposition~\ref{pro:single_k=1})}
Assertion (i) can be shown by rewriting the throughput as follows:
\begin{equation}
\theta_i = Z_n^{-1} \sum_l \abc(i,l,n) \sigma^l = Z_n^{-1} \sum_l \abc(n - i + 1,l,n) \sigma^l = \theta_{n - i + 1}.
\end{equation}
To prove assertion(ii) we first show that $(-1)^i(\theta_{i+1} - \theta_i)$ is positive. That is,
\begin{align}
\nonumber (-1)^i(\theta_{i+1} - \theta_i) &= (-1)^i Z_n^{-1} \sum_l \left( a(i+1,l,n) - a(i,l,n)\right) \sigma^l\\
\label{eqn:proof_lemma_single_k=1} &=(-1)^i Z_n^{-1} \sum_{l = i+1}^{\lfloor n/2 \rfloor} \left( a(i+1,l,n) - a(i,l,n)\right)\sigma^l > 0,
\end{align}
where the inequality follows from Lemma~\ref{lem:single_basic}. Using~\eqref{eqn:proof_lemma_single_k=1}, Proposition~\ref{pro:single_k=1}(ii) follows from
\begin{align*}
(-1)^i(\theta_{i+1} - \theta_i) &= (-1)^i \Big( \theta_{i+1} - Z_n^{-1} \sum_l a(i,l,n) \sigma^l\Big)\\
&=(-1)^i \Big( \theta_{i+1} - Z_n^{-1} \sum_l (a(i,l-1,n-2) +  a(i,l,n-1) )\sigma^l\Big)\\
&>(-1)^i \Big( \theta_{i+1} - Z_n^{-1} \sum_l (a(i,l-1,n-2) + a(i+1,l,n-1) )\sigma^l\Big)\\
&=(-1)^i \Big( \theta_{i+1} - Z_n^{-1} \sum_l a(i+2,l,n)\sigma^l\Big)\\
&=(-1)^{i+1}(\theta_{i+2} - \theta_{i+1}). 
\end{align*}
This completes the proof.
\end{proof}

\bibliographystyle{abbrv}
\bibliography{Philosopher}

\end{document}